\documentclass[
aps,
pra,
notitlepage,
citeautoscript, superscriptaddress,
reprint
]{revtex4-2}
\usepackage[dvipsnames,usenames]{xcolor}
\usepackage{graphicx}
\usepackage{graphicx}
\usepackage{dcolumn}
\usepackage{bm}
\usepackage{mathrsfs,natbib}
\usepackage[title]{appendix}
\usepackage{graphicx,epsf,amssymb,amsbsy,amsfonts,amssymb,amsmath}
\usepackage[colorlinks=true]{hyperref}
\hypersetup{
    unicode=false,          
    pdftoolbar=true,        
    pdfmenubar=true,        
    pdffitwindow=false,     
    pdfstartview={FitH},    
    pdftitle={},    
    pdfauthor={},     
    pdfsubject={},   
    pdfcreator={},   
    pdfproducer={}, 
    pdfkeywords={} {} {}, 
    pdfnewwindow=true,      
    colorlinks=true,       
    linkcolor=magenta, 
    citecolor=blue,        
    filecolor=magenta,      
    urlcolor=blue           
}
\usepackage{slashed}
\usepackage{comment}
\usepackage[dvipsnames]{xcolor}
\usepackage{color,soul}
\usepackage{simplewick}

\newcommand{\half}{\frac{1}{2}}
\newcommand{\be}{\begin{equation}}
\newcommand{\ee}{\end{equation}}
\newcommand\beq{\begin{eqnarray}}
\newcommand\eeq{\end{eqnarray}} 
\newcommand\eqn[1]{\label{eq:#1}} 
\newcommand\eq[1]{Eq.\ (\ref{eq:#1})}

\newcommand{\CO}{{\cal O}}

\newcommand{\CL}{{\cal L}}

\newcommand\vev[1]{\langle #1 \rangle}

\newcommand{\mybar}[1]

\newlength{\backup}

\begin{document}

\title{Hamiltonian framework for Chiral Gauge Theories on a Disk Boundary}
\author{Srimoyee Sen}%
\email{srimoyee08@gmail.com}
\affiliation{Department of Physics and Astronomy, Iowa State University, Ames, Iowa 50011, USA}%

\date{\today}

\begin{abstract}

I propose a Hamiltonian framework for chiral gauge theories (CGT) based on an existing Euclidean formulation, that uses 2n dimensional chiral fermions on the boundary of a 2n+1 dimensional disk. In the original Euclidean formulation,
boundary gauge fields were extended into the bulk using $2n + 1$
dimensional gauge field equations of motion (EOM). This creates a bottleneck for constructing a Hamiltonian. I present an alternate proposal for extending the gauge fields into
the bulk that is compatible with both a Hamiltonian framework and a Euclidean path integral. Applying this to Abelian gauge fields produces exact expressions of interior fields as a functional of the boundary fields, which can be directly used in a Hamiltonian formulation. 
Euclidean analysis of the new gauge field extension shows that it can preserve the non-perturbative
content of the original construction, including the behavior of topological charge, associated chiral
fermion zero modes and an absence of the strong CP problem when applied to the Standard Model.
This construction opens up a route to a Hamiltonian treatment and future
quantum simulation of CGTs on a disk boundary.

\end{abstract}
\maketitle
\section{introduction}
Despite the wild success of the Standard Model (SM) of particle physics, its non-perturbative definition remains out of our reach. In attempted lattice definitions, the hurdle arises from a conflict between the theory's chiral nature, where the weak sector discriminates between the left and right-handed fermions and the discrete nature of space-time which interferes with the conventional notion of chirality. In fact, the challenge applies to a class of theories known as chiral gauge theories which include the SM. 
Attempts in the past \cite{Luscher:1998du, Kadoh:2007wz, Kikukawa:2017ngf, Bock:1997ks} and the present have employed several techniques to tackle this problem with varying degrees of success. In fact, there has been considerable recent work using bosonization, Villain formulation and symmetry disentanglers \cite{Berkowitz:2023pnz, Baig:2026zgb, Lu:2026itw, Seifnashri:2026ema, Thorngren:2026ydw} and using symmetric mass generation (SMG) \cite{Wen:2013ppa, You:2014sqa, You:2014vea, You:2014oaa, Wang:2018ugf, Catterall:2020fep, Wang:2022ucy, Razamat:2020kyf, Nguyen:2024wck, Ayyar:2015lrd, Ayyar:2014eua, Catterall:2023nww}. This work concerns itself with a recent proposal to formulate CGT where the $2n$ dimensional target theory lives on the boundary of a $2n+1$ dimensional disk \cite{Kaplan:2023pxd, Kaplan:2023pvd}. The construction falls within a broader approach that uses an extra dimension of space to achieve its goal and I will refer to this approach as extra-dimension-approach (EDA) \cite{Kaplan:1992bt, Grabowska:2015qpk, Shamir:1993zy, Dang:2026pkb, Dang:2026xng, Kroth:2025hba, Araki:2026mqi, Aoki:2024bwx, Aoki:2022cwg, Aoki:2022aez}. The EDA approach in its current form is engineered for a Euclidean lattice and its potential Hamiltonian formulation suffers from a major identifiable  roadblock. This paper provides a strategy to overcome this problem for the disk construction, potentially clearing the path for a Hamiltonian treatment.

There are two distinct issues that plague lattice CGT. The first is the question of formulation. Plainly speaking, this means that most known discretizations of CGTs and all attempted discretizations of the SM in the past have failed to approach the target theory in the continuum limit. This ties to the Nielsen-Ninomiya theorem \cite{Karsten:1981gd, Nielsen:1980rz, Nielsen:1981hk, Nielsen:1981xu}, which forbids the realization of unpaired Weyl fermions in a finite $2n$ dimensional lattice. This theorem was eventually circumvented \cite{Kaplan:2023pvd} by the $2n+1$ dimensional disk proposal thereby enabling a possible framework for a formulation. In the disk formulation, unpaired $2n$ dimensional Weyl fermions are realized on the boundary of a disk and $2n$ dimensional boundary gauge fields are extended into the $(2n+1)^{\text{th}}$ dimension using $2n+1$ dimensional equation of motion (EOM), which I will refer to as an EOM flow. It belongs to a class of flows referred to as annealing flow, discussed in \cite{Kaplan:2024ezz}. 
The second issue concerning CGT is a question of computation with finite resources. It turns out that even when a formulation exists, Euclidean lattice Monte-Carlo will suffer from a sign problem \cite{Singh:2025sye}, making actual extraction of observables at strong coupling extremely difficult if not impossible. A Hamiltonian formulation could in principle circumvent this problem through quantum simulation. Thus, a Hamiltonian formulation of the disk construction is highly desirable. Note that, on the same day that the first version of this manuscript was posted, a paper extending Lüscher’s Abelian Euclidean path integral technique to non-Abelian theories appeared \cite{Craig:2026soq}.

There is an additional reason for seeking a Hamiltonian formulation. It relates to some of the striking features of the disk construction that were discovered when \cite{golterman2024conserved} surmised that the EDA approach had a potential flaw. \cite{golterman2024conserved} noted that when the disk construction is applied to the SM, in particular its QCD(quantum chromodynamics) subsector, boundary instanton (or anti-instanton) configurations of the gauge fields will produce singularities in the interior of the disk. 
Each of these defects (boundary instanton, anti-instanton and interior singularity) will carry localized chiral fermion zero modes, and the interior zero modes could turn the $\eta'$ particle into a Goldstone Boson (GB).
 However, \cite{Kaplan:2024ezz} showed  that this is not inevitable and illustrated how this can be avoided by working within a 't Hooft instanton gas model \cite{t1976symmetry}. The calculation in \cite{Kaplan:2024ezz} illustrated that under  annealing flow, where boundary instanton anti-instanton pairs are annihilated as a function of decreasing radial distance, interior singularities exist only when the net instanton number or topological charge on the boundary is nonzero: i.e. no interior singularities exist when the boundary topological charge is zero. This prevents the $\eta'$ from becoming a Goldstone mode \cite{Kaplan:2024ezz}. 
 In addition, \cite{Kaplan:2024ezz} argued that the behavior of instantons, the interior singularities and the associated chiral fermion zero modes turn the $\theta$-angle unphysical which eliminates the strong CP problem from the disk construction\cite{Kaplan:2024ezz}, \cite{t1976symmetry} . While this is quite remarkable within the context of the Euclidean action formulation, a pathway to constructing an explicit Hamiltonian will be extremely beneficial in understanding how the absence of the Euclidean strong CP problem on the disk connects to real time physics.

Even though a Hamiltonian formulation is desirable for the above reasons, the existing disk construction does not admit a natural Hamiltonian. The obstruction comes from the specific prescription of the annealing flow which uses the EOM. In this prescription the gauge field at a certain location in the extra dimension at a specific instant in time becomes dependent on boundary gauge fields at all other instants in time. Coupling bulk fermions to past and future boundary gauge fields is unsuitable for a Hamiltonian framework which creates a bottleneck for a Hamiltonian treatment of the disk. In fact, something very similar holds for another EDA based construction \cite{Grabowska:2015qpk, Dang:2026xng} which uses a slab manifold as opposed to a disk where $2n$ dimensional dynamical gauge fields are extended into the extra dimension using gradient flow.  

In this paper, I propose a prescription that overcomes this problem by constructing an instantaneous gauge field extension in the bulk. The prescription is suited for the disk construction and does not apply to the slab \cite{Grabowska:2015qpk}. Instead of solving a $2n+1$ dimensional EOM it solves a $2n$ dimensional EOM excluding time, at every time slice. This allows me to write interior gauge fields as an instantaneous functional of the boundary gauge fields. I solve for these functionals explicitly for Abelian gauge fields. The results can be directly used in a Hamiltonian formulation of Abelian CGTs and for the Abelian sector of non-Abelian CGTs in any dimension. While the framework applies to non-Abelian gauge fields, there are no analogous analytical expressions. In any case, for a lattice application the functionals will have to be obtained numerically. 
The proposed prescription is designed to eliminate bulk gauge field singularity 
when the net topological charge on the boundary is zero in congruence with \cite{Kaplan:2024ezz}. That it does so is checked explicitly in this paper for $n=1$ Abelian gauge fields. In addition, in the presence of net topological charge on the boundary, the singularity is pushed to the temporal boundary of the manifold. This too is established analytically in the case of $n=1$. For the case of $n=2$, i.e. 4 dimensional non-Abelian CGT on the boundary of a five dimensional disk, there is no topological obstruction to such a smooth extension and the $n=1$ calculation demonstrates how such smooth solutions can exist. Also, an explicit perturbative calculation following \cite{Luscher:2010iy} supports this conclusion in a weak coupling regime. Therefore, the paper tackles two problems: firstly it constructs a prescription for the gauge field that enables a Hamiltonian formulation, second the prescription retains the physics of $\eta'$ and the absence of strong CP problem seen in \cite{Kaplan:2024ezz}. 

The organization of the paper is as follows. In sec. \ref{sec2} I will review the results of \cite{Kaplan:2023pxd, Kaplan:2023pvd, Kaplan:2024ezz} and the calculations therein relating to the mass of $\eta'$ and the physics of the $\theta$ angle. Readers familiar with these results can skip to sec. \ref{sec3} where I introduce the new prescription for the gauge fields in the interior that is suitable for both a Euclidean action framework and a Hamiltonian. In sec. \ref{sec4}, I will discuss instanton physics in the new prescription and show why the behavior of topological charge follows the desired behavior in $n=1$ with a calculation and that the same is expected for $n=2$. Thus, 
$\eta'$ and the $\theta$ angle physics are expected to remain unchanged, i.e. the $\eta'$ avoids becoming a GB in the new prescription and the $\theta$ angle remains unphysical exactly as in \cite{Kaplan:2024ezz}.

\section{Euclidean action formulation with annealing flow}
\label{sec2}
To illustrate the disk construction proposed in \cite{Kaplan:2023pvd, Kaplan:2023pxd, Kaplan:2023pxd}, I begin with $n=2$ or five dimensional space-time in Cartesian coordinates given by $(x_t, x, y , w, z)$ or $(0, x, y , w, z)$. I will use $0$ and $x_t$ interchangeably, the former mostly as a subscript or superscript for a gauge field component and the latter to denote time coordinates. The manifold considered in \cite{Kaplan:2023pvd, Kaplan:2023pxd, Kaplan:2023pxd} corresponds to $\mathbb{T}_1\times \mathbb{R}_2\times \mathbb{T}_2$ where the Euclidean time $x_t$ and the spatial directions $w, z$ correspond to a three-torus whereas $x, y$ are coordinates on $\mathbb{R}_2$. Eventually, if we pursue a real time description, we will have to allow for $x_t$ to describe Minkowski time and it will not be compactified to a torus. 

One of the primary ingredients of the construction in \cite{Kaplan:2023pvd, Kaplan:2023pxd, Kaplan:2023pxd} is unpaired Weyl fermions. To engineer such Weyl fermions one begins with a massive Dirac fermion in five dimensions. In the continuum picture, one introduces a mass defect in $\mathbb{R}_2$, i.e. the Dirac mass is positive within a radius $R$ in cylindrical coordinates in $\mathbb{R}_2$ and then turns negative outside. The long wavelength spectrum of such a Dirac operator exhibits a four dimensional Weyl fermion of a fixed chirality localized at radius $R$, i.e. the boundary of the disk which is a manifold given by $\mathbb{T}_1\times \mathbb{S}_1\times \mathbb{T}_2$. An opposite chirality Weyl fermion can be obtained by simply flipping the sign of the mass defect. 
The same can be arranged on the lattice as shown in \cite{Kaplan:2023pvd, Dang:2026pkb} by using a Wilson fermion in five dimensions with an analogous mass defect and appropriate Wilson parameter. With a mass defect in the form of a disk, we switch to cylindrical coordinates $(x_t, r, \phi, w, z)$ where $r=\sqrt{x^2+y^2}$ and $\phi=\tan^{-1}(y/x)$.
Once the fermion content of the theory has been engineered this way, one simply turns on gauge coupling of the fermions and introduces a four dimensional gauge action for the gauge field in the path integral. The fluctuating gauge fields live on the boundary of the disk i.e. at $r=R$ and are themselves four dimensional, i.e. have $(0, \phi, w, z)$ components. 
A prescription is needed to couple a four dimensional gauge field to five dimensional fermions. The proposal in \cite{Kaplan:2024ezz} prescribes solving a five dimensional gauge field EOM (Yang-Mills for non-Abelian and Maxwell for Abelian) equation in the interior, i.e. $r<R$ for every four dimensional gauge field configuration on the boundary $r=R$. Note that this is one of the possible ways to extend the gauge fields into the bulk. There could in principle be other prescriptions as long as one maintains four dimensional gauge invariance of the target theory and obtains non-singular or smooth gauge fields in the interior when the boundary gauge field does not include nonzero net instanton number. E.g. a radially independent gauge field configuration preserves four dimensional gauge invariance. But it will lead to singularities at $r=0$ and is hence disfavored.\\ 

\noindent
{\bf Additional symmetry, $\eta'$ and the strong CP problem:} Beyond those mentioned above, there are additional considerations regarding the extension of the gauge field for $r<R$. To understand this, we need to focus on the role of instantons uncovered by \cite{Kaplan:2024ezz} in this construction. The discussion started with an observation by \cite{golterman2024conserved}, which noted that ABJ anomalies of the four dimensional target theory are tied to exact global symmetries of the five dimensional parent theory. \footnote{This also holds in $n=1$.}  \cite{golterman2024conserved} pointed out that this could result in a low energy theory whose phenomenology might deviate from the desired phenomenology of the target theory. In the case of the standard model, the concern arises within its quantum chromodynamics (QCD) subsector and can be illustrated simply using 1-flavor QCD. To engineer 1-flavor QCD on the boundary of the disk, one begins with two massive Dirac fermions in five dimensions and gives them opposite mass defects such that one produces a positive and the other a negative chirality Weyl fermion at $r=R$ which are then gauged using $SU(3)$ gauge fields. From the perspective of the five dimensional theory, each of the two Dirac flavors enjoys an exact $U(1)$ global symmetry. The two can be combined to construct the axial and the fermion number symmetries of the lower-dimensional target theory. From the point of view of the higher dimensional theory, both symmetries are exact. In purely four dimensional QCD, we know that the axial symmetry suffers from the ABJ anomaly. This in fact is responsible for the large mass of the $\eta'$, leading $1$-flavor QCD to be completely gapped even with a massless quark. In the case of the disk construction, the worry is that the exact symmetry of the parent theory, which is tied to the axial transformation of the target theory, may turn $\eta'$ into a Goldstone Boson upon chiral symmetry breaking. \cite{Kaplan:2024ezz} found a resolution to this. It noted that while the exact symmetry identified by \cite{golterman2024conserved} indeed exists, a GB $\eta'$ is prevented by the annealing flow of the gauge field and the resulting behavior of instantons. 

In the five dimensional disk construction with annealing flow, when boundary gauge configuration contains $n_i$
instantons and $\bar{n}_i$ anti-instantons, the interior contains $|n_i-\bar{n}_i|$ gauge singularities. Each of these defects hosts chiral fermion zero modes in their core, $q$ being the boundary fermion zero mode and $Q$ the bulk or interior fermion zero mode. The physical $\eta'$ is constructed out of the boundary $q$, quark fields. \cite{Kaplan:2024ezz} in fact modified the manifold from the disk to an annulus to accommodate the singularities on the inner boundary of the annulus. Following \cite{t1976symmetry}, \cite{Kaplan:2024ezz} summed up the instantons, anti-instantons and the singularities to write an effective partition function contribution for the $q$ and $Q$ fields

\beq
e^{-\tilde{S}_{\text{int}}}&\equiv & Z_\text{topo}\nonumber\\&=&\sum_{n_{\text{i}},\bar n_{\text{i}}}  \frac{(V \CO)^{n_\text{i}}}{n_{\text{i}}!} \frac{(V\bar \CO)^{\bar n_{\text{i}}}}{\bar n_{\text{i}}!} \left[ (V'  X)^{(n_{\text{i}}-\bar n_{\text{i}})} \Theta(n_{\text{i}}-\bar n_{\text{i}})\right.  \nonumber\\
&&\left. +(V'  \bar X)^{(\bar n_{\text{i}}-n_{\text{i}})} \Theta(\bar n_{\text{i}}-n_{\text{i}}) +\delta_{n_{\text{i}},\bar n_{\text{i}}}\right]\cr &&\cr 
&=& Z_1+Z_2+Z_3
\eqn{isum}
\eeq
where $V$ and $V'$ are the four volumes of the outer and inner boundary of the annulus, \beq
\Theta(n_{\text{i}}) = \begin{cases} 1 & n_{\text{i}}> 0\\ 0 & n_{\text{i}}\le 0 \end{cases}\ 
\eeq
and 
\beq
&&\CO =  \Lambda\int_V \frac{d^4x}{V}\,  \bar q_{R} q_{L} \ ,\quad
\bar \CO=  \Lambda\int_V \frac{d^4x}{V}\,  \bar q_{L} q_{R} \ ,\quad\nonumber\\
&&
X = \Lambda\int_{V'} \frac{d^4y}{V'}\,  \bar Q_{R} Q_{L} \ ,\quad\bar X =  \Lambda\int_{V'} \frac{d^4y}{V'}\,  \bar Q_{L} Q_{R} \ 
\eqn{ops}
\eeq
and $\Lambda$ is a QCD mass scale that cannot be computed within the model. 
Since the $\eta'$ spectrum in this construction arises from correlation functions of boundary fermion fields, i.e. $q$ fields, to arrive at the $\eta'$ EFT we simply need to focus on $Z_3$ in \eq{isum}. 
From here, one matches onto the EFT for the $\eta'$, while introducing a boundary quark mass of $m_q$ and a $\theta$ angle, with

\beq
 \bar  q_R  q_L(x) &\to &\frac{\Sigma}{2} e^{i\eta'(x)/f}\ ,\qquad 
 \bar  q_L  q_R(x)\to \frac{\Sigma}{2}e^{-i\eta'(x)/f}\nonumber\\
 \frac{\Lambda \Sigma}{ f^2} &=& M_{\eta'}^2\ ,\qquad \qquad\quad\frac{m_q \Sigma}{ f^2} = M_{q}^2
\eqn{qsub}
\eeq
 where $\Sigma=\vev{\bar q q}$ is the chiral condensate. 
The corresponding $\eta'$ EFT is given by 
\beq
\widetilde \CL_{\eta'} &=&
 \half  \partial_\mu\eta' \partial_\mu\eta'(x)   +\nonumber\\
 &+&M_{\eta'}^2 f^2\left[ -1 + \frac{\left(\eta'(x) - \overline{\eta'} \right)^2}{2f^2} + O(\eta'^4 )\right]+\nonumber\\
 &&- M_{q}^2 f^2 \cos\left(\frac{\eta'}{f}+\theta\right)
\ 
\eqn{Leta}
\eeq
where \beq
\overline{\eta'} \equiv \int \frac{d^4x}{V} \, \eta'(x)\ .
\eeq 
Thus, we can see that in the limit of $m_q, M_q=0$, there exists a shift symmetry in the effective Lagrangian
\beq
\eta'\rightarrow \eta' +\delta\eta'
\label{shift}
\eeq
where $\delta\eta'$ is a constant. This is how the exact ``axial symmetry" of the parent theory gets realized in the lower dimensional target theory without realizing a GB. 
It results in gaplessness of the zero momentum mode of the $\eta'$ which is a global excitation. However, every other momentum mode of the $\eta'$ experiences a gap of $M_{\eta'}$ matching the expected phenomenology of QCD. In addition, the zero momentum mode of the $\eta'$ allows one to rotate away the $\theta$ angle which makes it unphysical as elaborated in \cite{Kaplan:2024ezz}.

\subsection{Gauge field functional and challenges to constructing a Hamiltonian}
A Hamiltonian for the free fermion sector of the disk construction can easily be obtained. It is the gauge fields where subtleties arise. In general, for $r<R$, any component of the gauge field  will have to be expressed as a functional of the boundary gauge fields.

For the $2n+1$ dimensional EOM flow, to obtain the interior gauge field one solves the equation
\beq
D_A \bar{F}_{AB}=0
\label{eom}
\eeq
where the capitalized indices $A, B$ go over all five dimensional space-time coordinates and $D_A$ is the covariant derivative.
The equation is solved subject to the boundary condition at $r=R$. 
 $\bar{A}_A$ gives gauge fields for all $r$ with $\bar{A}_r(r=R)\equiv 0$ and $\bar{A}_{\mu}(r=R)\equiv A_{\mu}$  where  $A_\mu$ is the boundary gauge field with $\mu$ going over time and all spatial indices except $r$, $\bar{F}_{AB}$ is the field strength for gauge field $\bar{A}_A$. Thus, Eq. \ref{eom}, upon solving will produce a functional of $A(x_t, R, \phi, w, z)$ for $\bar{A}$ at every $r$. I don't solve for this functional for the EOM flow in this paper. However, for a potential Hamiltonian formulation following Eq. \ref{eom} one will need such a functional written in terms of gauge links.  
\footnote{Note that the above constraint on the interior gauge field can easily be expressed in the form of a completely local action formulation where the constraint is implemented using a Lagrange multiplier field at every interior space-time points. 
The novelty is that such a local action leads to the precise non-local action in \eq{Leta} that allows for the mostly gapped spectrum of $\eta'$ and can lead to the elimination of the strong CP problem from this construction.}

 To see why a simple Hamiltonian treatment for the EOM fails, it is useful to consider the boundary $r=R$ ansatz of 
 \beq
 A_w=a_w(x_t),\, A_{0,z,\phi, r}=0.
 \eeq
 $a_w(x_t)$ is a function of $x_t$ only and does not depend on $\phi, w, z$. 
 The EOM of Eq. \ref{eom} is satisfied by the solution
 \beq
 \bar{A}_w(x_t^{0}, r)=\int a_w(\tilde{x}_t)\frac{I_0(\omega r)}{I_0(\omega R)}e^{i\omega(\tilde{x}_t-x_t^0)}d\tilde{x}_t\, \frac{d\omega}{2\pi}
 \eeq
 where I have assumed Euclidean time. 
 Thus we see that $\bar{A}_w(x_t^{0}, r)$ depends on $\bar{A}_w(x_t, r=R)$ or $A_w(x_t)$ at all $x_t$ including $x_t=x_t^0$. This is not suitable for a Hamiltonian formulation if our strategy is to express interior gauge fields as functionals of the boundary gauge field.

\section{The new proposal and the flow calculation}
\label{sec3}
 Here I propose a gauge field extension into the interior to make it suitable for a Hamiltonian framework. The new prescription is simply solving a $2n+1$ dimensional Yang-Mills or Maxwell's EOM with the time derivatives omitted for the spatial gauge field components:
\beq
D_i \bar{F}_{iA}=0
\label{eqmod}
\eeq
where $i$ goes over only spatial indices $(r, \phi, w, z)$ and `A' over all five coordinates as before $(0, r, \phi, w, z)$. For the spatial gauge fields, \ref{eqmod} is simply $2n$ dimensional Yang-Mills or Maxwell's EOM in $(r, \phi, w, z)$.
Eq. \ref{eqmod} is valid at every instant in time $x_t$ and is designed to produce a functional for the interior gauge fields that is instantaneous.

The task is to solve Eq. \ref{eqmod} in order to understand the flow. For non-Abelian gauge fields, an analytic solution for the interior gauge fields in terms of the generic boundary gauge field data is not possible and should be tackled numerically. However, \ref{eqmod} can be solved exactly for Abelian gauge fields. Thus, the solution functionals can be used in Hamiltonian Abelian CGT formulations. In addition, the Abelian solutions can capture non-Abelian results in the weak coupling limit or when the nonliner terms in Eq. \ref{eqmod} are small compared to the linear ones. This was first discussed in  \cite{Luscher:2010iy}, which proposed a path to solving Wilson flow for non-Abelian gauge fields analytically in the small coupling limit. There, at leading order in the coupling, the relevant Wilson flow equation reduces to an Abelian gauge field equation. The non-Abelian corrections can be added systematically order by order in the coupling $g_{\text{YM}}$. I will solve the Abelianized equation for $n=2$. The $n=1$ results can simply be read off from $n=2$. The $n=1$ Abelian equations and their solution are very useful in revealing the behavior of lower dimensional boundary instantons without needing any approximation. 

The Abelian version of \ref{eqmod} is
\beq
\partial_i \bar{F}_{iA}=0
\eeq
with 
\beq
\bar{F}_{iA}=\partial_i \bar{A}_A-\partial_A \bar{A}_i.
\eeq

Let's focus on solving the equations for $A=r, \phi, w, z$. As we will see, the time component equations can be solved separately. I will write the equations in cylindrical coordinates explicitly.\\

\noindent
{\bf $\hat{r}$ component:}
\beq
\frac{1}{r^2}\partial_\phi\partial_r(r \bar{A}_\phi)-\frac{1}{r^2}\partial_\phi^2 \bar{A}_r&-&\left(\partial_z^2\bar{A}_r-\partial_z\partial_r \bar{A}_z\right)\nonumber\\
&&-\left(\partial_w^2 \bar{A}_r-\partial_w\partial_r \bar{A}_w\right)=0\nonumber\\
\label{ar}
\eeq

\noindent
{\bf $\hat{\phi}$ component:}
\beq
\frac{1}{r}\partial_\phi\partial_z \bar{A}_z&-&\partial_z^2 \bar{A}_\phi+\frac{1}{r}\partial_\phi\partial_w \bar{A}_w-\partial_w^2 \bar{A}_\phi\nonumber\\&-&\partial_r\left(\frac{1}{r}\partial_r(r \bar{A}_\phi)\right)+\partial_r\left(\frac{1}{r}\partial_\phi \bar{A}_r\right)=0\nonumber\\
\label{aphi}
\eeq

\noindent
{\bf $\hat{z}$ component:}
\beq
\frac{1}{r}\partial_r\left(r \partial_z \bar{A}_r\right)&-&\frac{1}{r}\partial_r\left(r \partial_r \bar{A}_z\right)+\frac{1}{r}\partial_\phi\partial_z \bar{A}_\phi\nonumber\\ &-&\frac{1}{r^2}\partial_\phi^2 \bar{A}_z+\partial_z\partial_w \bar{A}_w-\partial_w^2 \bar{A}_z=0\nonumber\\
\label{az}
\eeq

\noindent
{\bf $\hat{w}$ component:}
\beq
\frac{1}{r}\partial_r\left(r \partial_w \bar{A}_r\right)&-&\frac{1}{r}\partial_r\left(r \partial_r \bar{A}_w\right)+\frac{1}{r}\partial_\phi\partial_w \bar{A}_\phi\nonumber\\
&-&\frac{1}{r^2}\partial_\phi^2 \bar{A}_w+\partial_w\partial_z \bar{A}_z-\partial_z^2 \bar{A}_w=0\nonumber\\
\label{aw}
\eeq

Note that if we define the LHS of the above equations as, 
$\bar{E}_r, \bar{E}_\phi, \bar{E}_z, \bar{E}_w$, then the following identity holds for any $\bar{A}$ (and not just the solution to the above equations)
\beq
\partial_\phi \bar{E}_\phi+\partial_z(r \bar{E}_z)+\partial_w(r \bar{E}_w)+\partial_r(r \bar{E}_r)
\label{e2}
=0\eeq

In the discussion below I will solve the flow equation \ref{eqmod} subject to the boundary data $A_{z,w,\phi}$ and $A_r=\bar{A}(R)=0$ while fixing a gauge so as to obtain interior gauge fields as unique functionals of the boundary gauge fields. However, one can solve \ref{eqmod} without a gauge fixing in which case some of the interior gauge field components will be expressed as a functional of the boundary gauge fields and the remaining interior gauge field components.

Eq. \ref{eqmod} respects five dimensional gauge invariance. Since, we only care about four dimensional gauge invariance, i.e. in $x_t, \phi, w, z$, I could in principle fix the gauge to $\bar{A}_r=0$ to solve the above equations \cite{Dang:2026pkb, Dang:2026xng}. However, such a gauge does not lead to smooth solutions for all possible boundary data. To illustrate, I will go to Fourier space with momentum $p_w=0, p_z=0$ in $w, z$ and azimuthal wave number $m=1$ and consider boundary data
\beq
A_{w,z}(R)=0, A_\phi(R)=c e^{i(\phi)}.
\label{bcd}
\eeq
 Attempting to solve Eq. \ref{ar}, \ref{aphi}, \ref{aw} and \ref{az}, we see that the equations are trivially satisfied with radially independent copies of $\bar{A}_{w,z}$ and Eq. \ref{aphi} in conjunction with requiring a non-singular solution at $r=0$ forces
\beq
\bar{A}_\phi=\frac{c r}{R} e^{i\phi}.
\eeq
This however does not satisfy Eq. \ref{ar} unless $c=0$. Thus, arbitrary boundary data for the gauge fields $A_{z,w,\phi}$ are not consistent simultaneously with demanding that the solutions are non-singular at $r=0$ and $\bar{A}_r=0$.

To solve the equations Eq. \ref{ar}, \ref{aphi}, \ref{aw} and \ref{az}, while avoiding singularity, I will make the following gauge choice
\beq
\Delta \partial_i \bar{A}_i=0
\label{eqda}
\eeq
where $\Delta$ is the Laplacian in $(r, \phi, w, z)$.\\ 

The intermediate steps to solving Eq. \ref{ar}, \ref{aphi}, \ref{aw}, \ref{az} and Eq. \ref{eqda} are given in the appendix \ref{app1}. Below I only present the solutions in terms of the constants of integration (see Eq. \ref{afin1} and \ref{afin2}). The expressions below were obtained for $p\neq 0$ demanding non-singular solutions

\beq
&&\bar{A}_z=\left(\tilde{C}_1 I[m, p\, r]+ i p_z r \,C_3 I[m+1, p\, r]\right)\nonumber\\ &&\hspace{2 in}e^{i(m\phi+p_z z+p_w w)}\nonumber\\
&&\bar{A}_w=\left(\tilde{C}_2 I[m, p\, r]+ i p_w r \,C_3 I[m+1, p\, r]\right)\nonumber\\&&\hspace{2 in}e^{i(m\phi+p_z z+p_w w)}.\nonumber\\
\label{asol}
\eeq
\beq
&&\bar{A}_r=\frac{1}{2}\left[C_+ I[m+1, p\, r]+C_- I[m-1, p\, r]\right.\nonumber\\
&&\left.+ C_3 r \left(\frac{d}{dr}(I[m+1, p \, r])+ \frac{d}{dr}(I[m-1, p \, r])\right)\right]\nonumber\\&& \hspace{2 in}e^{i(m\phi+p_w w+p_z z)}\nonumber\\\nonumber\\
&&\bar{A}_\phi=\frac{1}{2i}\left[C_ + I[m+1, p\, r]-C_- I[m-1, p\, r]\right.\nonumber\\
&&\left.+ C_3 r \left(\frac{d}{dr}(I[m+1, p \, r])- \frac{d}{dr}(I[m-1, p \, r])\right)\right]\nonumber\\&& \hspace{2 in}e^{i(m\phi+p_w w+p_z z)}.\nonumber\\
\label{asol2}
\eeq
Here, $I$ denotes the modified Bessel functions of the first kind.
Thus, I have five unknown constants of integration. The gauge condition itself provides another equation
\beq
\frac{p}{2}(C_++C_-)+C_3 p(m-1)+i p_z \tilde{C}_1+i p_w \tilde{C}_2=0.\nonumber\\
\label{g2}
\eeq

The five unknown integration constants can be solved using the boundary values of the components $\bar{A}_z(R)=A_z, \bar{A}_w(R)=A_w, \bar{A}_\phi(R)=A_\phi$ as well as $\bar{A}_r(R)=A_r=0$ and the gauge condition equation \ref{g2}.

For $p=0$, the expressions are:
\beq
\bar{A}_w&=&\begin{cases}A_w\left(\frac{r}{R}\right)^{|m|} \,\, \text{for}\,\,m\neq 0\\
A_w\,\, \,\,\,\,\,\,\,\,\,\,\,\,\,\,\,\,\,\text{for}\,\,m= 0\end{cases}\nonumber\\
\bar{A}_z&=&\begin{cases}A_z\left(\frac{r}{R}\right)^{|m|} \,\, \text{for}\,\,m\neq 0\\
A_z\,\, \,\,\,\,\,\,\,\,\,\,\,\,\,\,\,\,\,\text{for}\,\,m= 0\end{cases}\nonumber\\
\bar{A}_r&=&\begin{cases}A_\phi\frac{|m|(|m|+2)}{2i m}\left(\left(\frac{r}{R}\right)^{|m|-1}-\left(\frac{r}{R}\right)^{|m|+1}\right) \,\, \text{for}\,\,m\neq 0\\
0\,\, \,\,\,\,\,\,\,\,\,\,\,\,\,\,\,\,\,\text{for}\,\,m= 0
\end{cases}\nonumber\\
\bar{A}_{\phi}&=&\begin{cases}\frac{A_\phi}{2}\left((|m|+2)\left(\frac{r}{R}\right)^{|m|-1}-(|m|)\left(\frac{r}{R}\right)^{|m|+1}\right)\\
\hspace{2.3 in}\text{for}\,\,m\neq 0\\\\
A_\phi \left(\frac{r}{R}\right)\,\hspace{1.8in}\text{for}\,\,m= 0\nonumber\\
\end{cases}
\eeq

I can now consider the equation for $\bar{A}_0$, given by 

\beq
&&\frac{1}{r}\partial_r\left(r\, \partial_r \bar{A}_0\right)+\partial_w^2 \bar{A}_0+\partial_z^2 \bar{A}_0+\frac{1}{r^2}\partial_\phi^2 \bar{A}_0\nonumber\\
&&=\partial_0\left(\partial_w \bar{A}_w + \partial_z \bar{A}_z+\frac{1}{r}\partial_\phi \bar{A}_\phi+\frac{1}{r}\partial_r(r \bar{A}_r)\right)
\eeq
which I can solve using Green's function for the equation
\beq
&&\frac{1}{r}\partial_r\left(r\, \partial_r G\right)+\partial_w^2 G+\partial_z^2 G+\frac{1}{r^2}\partial_\phi^2 G=\frac{\delta(r-\rho)}{r}\nonumber\\
\eeq
Fourier transforming to azimuthal wave number and momenta $(m, p_w, p_z)$ I get
\beq
\frac{1}{r}\partial_r\left(r\, \partial_r G\right)-p^2 G-\frac{m^2}{r^2} G=\frac{\delta(r-\rho)}{r}
\label{G}
\eeq
with
\beq
&&G(r,\rho)\nonumber\\
&&=\begin{cases}
 -I[m, p r]\left(K[m, p\rho]-\frac{K[m, p R]}{I[m,p R]}I[m,p\rho]\right)\nonumber\\
\hspace{2.1 in}\,\,\,\,\text{for}\,\,\,\,r<\rho\\\nonumber\\
-I[m,p\rho]\left(K[m, p r]-\frac{K[m, p R]}{I[m, p R]}I[m, p r]\right)\nonumber\\
\hspace{2.1 in}\,\,\,\,\text{for}\,\,\,\,r>\rho
\end{cases}\nonumber\\
\eeq
with $p\neq 0$. For $p=0$ the expressions are
\beq
&&G(r,\rho)\nonumber\\\nonumber\\
&=&\begin{cases}
-\frac{1}{2|m|}\left(\left(\frac{r}{\rho}\right)^{|m|}-\left(\frac{r\rho}{R^2}\right)^{|m|}\right) \,\, \text{for}\,\, m\neq 0, r<\rho\nonumber\\\nonumber\\-\frac{1}{2|m|}\left(\left(\frac{\rho}{r}\right)^{|m|}-\left(\frac{r\rho}{R^2}\right)^{|m|}\right) \,\,\, \,\text{for}\,\, m\neq 0, r>\rho\nonumber\\\nonumber\\
\log\left(\frac{\rho}{R}\right)\,\, \hspace{1 in}\text{for}\,\, m= 0, r<\rho
\nonumber\\\nonumber\\
\log\left(\frac{r}{R}\right)\,\, \hspace{1 in}\text{for}\,\, m= 0, r>\rho.
\end{cases}
\eeq
I define $\partial_{0}(\partial_i \bar{A}_i)=\partial_{x_t}(\partial_i \bar{A}_i)\equiv f(x_t, r,\phi,w,z)$.
Thus, the gauge field $\bar{A}_0$, for $p\neq 0$ is given by
\beq
&&\bar{A}_0(x_t,r, m, p_w, p_z)=A_0(x_t, R, m, p_w,p_z)\frac{I[m,p r]}{I[m, p R]}\nonumber\\
&&+\int_0^R G(r,\rho)f(x_t, \rho, m, p_w,p_z)\rho d\rho.
\label{a0}
\eeq
Analogous expression for $p=0$ can easily be obtained using the provided Green's function.

This provides $\bar{A}$ gauge fields for all $(r, m, p_w, p_z)$ for a specific $x_t$ written as functionals of the boundary fields. And the solutions are explicitly non-singular in the interior.

\subsection{Why this works only for the disk and not the slab}

This subsection is a short detour to an earlier EDA proposal constructed in \cite{Grabowska:2015qpk}. Readers focused on the disk construction can skip to the next section. The earlier EDA proposal \cite{Grabowska:2015qpk} too was aimed at Euclidean action formulation.  The five dimensional manifold in that case was a slab with the following coordinates: $s$ is the $(2n+1)^{\text{th}}$ or the extra dimension, whereas $x_t, v, w, z$ are the $2n$ dimensions (with $n=2$) for the target theory, $x_t$ being the time direction. The slab contains a near domain wall at $s=0$ and an anti-wall (or far wall) at $s=L$ with Weyl fermions of desired chiralities located on the near wall and mirror fermions of opposite chiralities on the far wall. To construct a chiral gauge theory \cite{Grabowska:2015qpk} proposed turning on four dimensional gauge fields at $s=0$ which are then damped using gradient flow in the extra dimension, so they decouple from the mirror wall. However, the gradient flow equation involves time derivatives of gauge fields in a way that makes extra dimension gauge fields functional of the $s=0$ gauge fields at all instants in time, posing the same problem as discussed earlier. 

The question is whether the prescription in Eq. \ref{eqmod} when applied in Cartesian coordinates can work in the case of the slab. The answer to this question appears to be negative.
 The reason for this can be understood by considering purely time dependent gauge fields. Let's first consider this in the disk case. 
A boundary condition of $A_z=F(x_t)$ with all other gauge fields being zero is solved by $\bar{A}_z(r)=\bar{A}_z(R)=A_z=F(x_t)$. Thus, the configuration $\bar{A}_z$ does not damp to zero in the interior and remains finite and smooth. This works perfectly fine with the disk construction since there are no massless mirror chiral fermions in the interior of the disk to decouple. However, an analogous boundary ansatz will cause the slab construction to fail. E.g. consider the slab geometry with Eq. \ref{eqmod} in Cartesian coordinates $s, w, z, v$. Since the EOM is a second order equation, to solve it we need another boundary condition at $s=L$. It is quite clear that the field strengths must go to zero at $s=L$ to decouple the mirror fermions. Therefore, we can use this as a second boundary condition.
If we now consider $s=0$ gauge field ansatz of $A_z=F(x_t)$, we see that the only solution to the equations Eq. \ref{eqmod} is a constant $\bar{A}_z$ which implies that the boundary conditions at $s=0$ and $s=L$ are not compatible. In other words, for certain gauge fields at $s=0$, it is impossible to decouple the mirror fermions on the anti-wall at $s=L$. Stated another way, the requirement of decoupling the mirror fermions on the anti-wall, appears to forbid certain field strengths from getting realized on the wall at $s=0$, thus  sabotaging the chiral gauge theory construction. The slab formulation may admit a Hamiltonian through a different procedure. However, the approaches so far including the one proposed in this paper do not apply.

\section{Instantons}
\label{sec4}
Eq. \ref{eqmod} is of course compatible with any boundary condition in $x_t$, including periodic and open. 
And, the equation is also compatible with both Minkowski and Euclidean spacetime. It is therefore interesting to compare Eq. \ref{eom} and Eq. \ref{eqmod} in Euclidean time with PBC in $x_t$ so that the boundary manifolds in both cases is $\mathbb{T}_1\times \mathbb{S}_1\times \mathbb{T}_{2(n-1)}$. With this, one can consider solving the two equations for the solid torus contained in this boundary. The comparison is especially interesting for boundary instanton configurations, which are the focus of this section. We expect single instanton boundary configurations to produce a singularity somewhere in the interior of the solid torus in both cases. For Eq. \ref{eom}, we know that under annealing flow the $(x_t, \phi, w, z)$ coordinate of the interior singularity will not correlate with the $(x_t, \phi, w, z)$ coordinate of the core of the boundary instanton. This means that the coordinates of the interior fermion zero mode are uncorrelated with those of the boundary fermion zero mode, which is reflected in  \eq{isum}. 
Eq. \ref{eqmod} with its associated gauge condition is designed to achieve analogous physics: for most $x_t$, including around where instanton field strength peaks, the interior fields are expected to be completely smooth and non-singular. This is because while a singularity is required somewhere in the interior, its location is not fixed. For every $x_t$, Eq. \ref{eqmod} is solved slice by slice in $x_t$ for the instantaneous configuration of the boundary gauge field which does not carry information of the total instanton number over the entire boundary space-time manifold. 
Thus, with the right gauge condition,  the second order equation Eq. \ref{eqmod} should produce perfectly smooth and continuous gauge fields in the interior for any allowed boundary gauge fields for almost all of $x_t$.  
It is only at the temporal boundaries, where we expect interior ($r<R$) singularities. I verify these expectations with an explicit calculation for Abelian CGTs in $1+1$ dimensions or $n=1$. For the non-Abelian case, exact calculation for all boundary gauge fields will involve numerical work beyond the scope of this paper. However, we expect such solutions to exist since there are no obstructions to continuing $r=R$ boundary gauge fields $A_{\phi, w, z}$ and $\bar{A}_r(R)=0$ at time slice $x_t$ to the interior $r<R$.

To verify the above, I first construct some configurations with nonzero instanton numbers on purely $2n$ dimensional manifolds and then assess their fate when placed on the boundary of the ($2n+1$) dimensional disk. As I construct these ansatze, I will prefer a gauge choice which sets $A_0=0$ in most of space-time, to align with the most common gauge choice in Hamiltonian formulation. However, the results I present don't depend on this choice. 
Here I will begin with a $2n$-dimensional torus with $n=1$ to build intuition and subsequently go to $n=2$ for the four dimensional instanton discussion. 

For the $n=1$ calculation, I will denote the spatial direction as $x_1$ and the Euclidean time as $x_t$. 
In the continuum, I can construct the simplest instanton configuration using the following ansatz (setting the coupling $e\rightarrow 1$) 
\beq
A_1= -\frac{2\pi x_t}{L_t L_1}, \,\,A_0=0
\label{a1}
\eeq
for $L_t-\epsilon \geq x_t\geq +\epsilon$ with $\epsilon\rightarrow 0^+$
where $L_t$ and $L_1$ are volumes of the two dimensions. The corresponding electric field can be found as 
\beq
E_1=\frac{2\pi}{L_t L_1}. 
\label{1din}
\eeq
Thus, the topological charge for this configuration is
\beq
\int dx_1 dt\,\, \left(\frac{E_1}{2\pi}\right)=1. 
\eeq
Note that the gauge configuration is not periodic in $x_t$ and instead is gauge equivalent to itself after one period. This is to say: 
\beq
A_1(L_t, x_1)&=& A_1(0,x_1)+\partial_1 \omega\nonumber\\
\omega &=&- \frac{2\pi x_1}{L_1}\nonumber\\
\label{omega}
\eeq
with $x_t=0, L_t$ identified. At the temporal boundary we need an $A_0$ gauge field of the form
\beq
A_0=\frac{2\pi x_1}{L_1}\tilde{\delta}(x_t), \int dx_t\tilde{\delta}(x_t)=1
\label{a0bdry}
\eeq
where $\tilde\delta(x_t)$ has support over  $L_t+\epsilon>x_t>L_t-\epsilon$ with the requirement that $\epsilon\rightarrow 0^+$ with $x_t=0$ and $L_t$ are identified.

The same can be understood on the lattice quite easily,   with $L_t$ sites in $x_t$ and $L_1$ sites in $x_1$ where $L_t$ and $L_1$ can only be integers. The gauge configuration can be represented simply as 
\beq
U_1(x_t, x_1)=\exp\left(-i \frac{2\pi x_t}{L_t L_1}\right)
\eeq
for $x_t=0, \cdots L_{t}-1$
and 
\beq
 U_0(x_t, x_1)=\begin {cases}1\hspace{.9 in} \text{for $x_t=0, \cdots L_t-2$}\\
 \exp(i\frac{2\pi x_1}{L_1})\hspace{0.4 in} \text{for $x_t=L_t-1$}.
 \end{cases}
\eeq
Computing the plaquettes, it is easy to see that the electric field equals the expression in Eq. \ref{1din} for each plaquette and we obtain a single instanton configuration this way. 

I can place this instanton on the boundary ($\mathbb{T}_1\times \mathbb{S}_1$) of the $2+1$ dimensional disk ($\mathbb{T}_1\times \mathbb{D}_2$) of radius $R$, by substituting $x_1 \rightarrow R\phi$ in the gauge field ansatz of Eq. \ref{a1}, and setting $L_1$ to $2\pi R$. I can now solve for the gauge field in the interior of the disk away from the temporal boundary at $x_t=0, L_t$ according to our flow prescription which leads to the equations

\beq
    \partial_r
    \left[
    \frac{1}{r}(\partial_r(r \bar{A}_\phi))
    \right]-\partial_r\left(\frac{1}{r}\partial_\phi \bar{A}_r\right)
    = 0,\nonumber\\
    \frac{1}{r^2}\partial_\phi\partial_r(r\bar{A}_\phi)-\frac{1}{r^2}\partial_\phi^2\bar{A}_r=0
    \label{eq:blk2}
\eeq
with the gauge condition
\beq
\Delta_2\left(\frac{1}{r}\partial_\phi\bar{A}_\phi+\frac{1}{r}\partial_r(r\bar{A}_r)\right)=0
\label{gc}
\eeq
and equation for $0^{\text{th}}$ component
\beq
\frac{1}{r}\partial_r(r\partial_r \bar{A}_0)+\frac{1}{r^2}\partial_\phi^2 \bar{A}_0=\partial_{0}\left(\frac{1}{r}\partial_\phi\bar{A}_\phi+\frac{1}{r}\partial_r(r \bar{A}_r)\right)\nonumber\\
\label{a0bar}
\eeq
where $\bar{A}$ is a function of $x_t, r, \phi$ and $\bar{A}_{0, r,\phi}(x_t, R,\phi)=A_{0, r,\phi}(x_t, \phi)$, $\Delta_2$ is the two-dimensional Laplacian in $r, \phi$ and Eq. \ref{gc} is the gauge condition analogous to Eq. \ref{eqda}. It is easy to see that $\bar{A}_r=0$ satisfies the equations with a general solution to Eq. \ref{eq:blk2} and Eq. \ref{gc} away from $r=0$  given by
\beq
\bar{A}_\phi=\frac{\kappa^{(1)}R}{r}+\frac{\kappa^{(2)} r}{R}
\eeq
where $\kappa^{(1)}+\kappa^{(2)}=-( x_t)/( R L_t)$. Demanding a non-singular solution at $r\rightarrow 0$ leads to $\kappa^{(1)}=0$ producing 
\beq
\bar{A}_\phi=\frac{\kappa^{(2)} r}{R}=-\frac{2\pi x_t r}{L_1 L_t R}=-\frac{ x_t r}{  R^2 L_t}.
\label{aphibar}
\eeq
Eq. \ref{a0bar} can then be solved with 
\beq
\bar{A}_0=0.
\label{a0b}
\eeq

This is one of the key ideas of this construction. Note that $\bar{A}_\phi, \bar{A}_0$ do not include any singularity for any $x_t$ away from the temporal boundary. Instead, the singularity is found specifically at $x_t=0$ or $L_t$, i.e. at the temporal boundary in the interior of the disk ($r<R$). The exact $r$ value and the shape of the discontinuity can not be obtained from the analysis of this paper, since the lattice calculation will involve solving the relevant equations for the gauge links themselves as opposed to vector potentials. However, here is an illustration of the singularity in the continuum gauge potential language by considering the boundary $A_0$ ansatz of Eq. \ref{a0bdry}. 
Note that this boundary data for the gauge potential $A_0$ is discontinuous in $\phi$. However, in terms of the gauge link variables $U_0(L_t, x_1)$, there is no discontinuity as
$U_0(L_t, x_1)=U_0(L_t, x_1+L_1)$ or $U_0(L_t, \phi)=U_0(L_t, \phi+2\pi)$. Here, while considering, $\mathbb{T}_1\times \mathbb{D}_2$, I have adopted the convention that gauge links without over-bar correspond to $r=R$ gauge links, whereas those with over-bar include all of $r$.  
The following solves equation \ref{a0bar} while satisfying the requisite boundary data of Eq. \ref{a0bdry}
\beq
\bar{A}_0= c_1 \frac{\log(r/r_0)}{\log(R/r_0)}\phi\,\tilde{\delta}(x_t)+   c_2 \phi\,\tilde{\delta}(x_t)
\eeq
with $c_1+c_2=1$.
However, the solution is singular at $r\rightarrow 0$ and furthermore for $c_1\neq 0$, the gauge links $\bar{U}_0(r, x_t, \phi)$ are multivalued as a function of $\phi$ for generic values of $r<R$. Thus, we see the interior singularity and discontinuity appear at $x_t=0, L_t$.

The instanton configuration of Eq. \ref{a1} is spread out. 
 However, it can be made localized in the time direction simply by replacing $x_t$ with $\Theta(x_t- x_t^{(0)})L_t$ where $L_t>x_t^{(0)}>0$ and $\Theta$ is a smooth approximation of a unit step function. We can also localize the instanton configuration in $x_1$ by scaling the ansatz in Eq. \ref{a1} with $L_1 f(x_1)/(\delta l)$ where $f(x_1)$ is a smooth approximation to a box function with support over $\delta l$. Note that the spread out or the uniform boundary configuration of Eq. \ref{a1} with topological charge $1$ solved Eq \ref{eq:blk2} and \ref{a0bar} with $\bar{A}_r=0$ which is compatible with the gauge choice of Eq. \ref{gc}. However, it is also compatible with a gauge choice of $\bar{A}_r=0$. In contrast, when the configuration gets localized in $x_1$ or $\phi$,  the gauge choice of Eq. \ref{gc} produces smooth and non-singular solution at $r=0$, but $\bar{A}_r=0$ does not. 

I consider an instanton anti-instanton pair separated in time. The simplest ansatz is the following where we have a temporally localized instanton-anti-instanton pair:
\beq
A_1= -\frac{2\pi \Theta(x_t- x_t^{(0)})\Theta( x_t^{(1)}-x_t)}{ L_1}, \,\,A_0=0
\label{inanin}
\eeq
where I assume $x_t^{(0)}$ and $x_t^{(1)}$ to be well separated and $x_t^{(1)}\gg x_t^{(0)}\gg 0$. The above ansatz holds for all $x_t$ including the temporal boundary. There is absolutely no discontinuity in $A_1$, i.e. it is single-valued in the entire $1+1$ dimensional manifold. Also, $A_0$ can be set to zero everywhere in the $1+1$ dimensional manifold as seen in Eq. \ref{inanin}. One can now solve for the interior gauge field using Eq. \ref{eq:blk2} and Eq. \ref{a0bar} where the solutions are smooth and finite everywhere given by
\beq
\bar{A}_\phi &=&-\frac{2\pi \Theta(x_t- x_t^{(0)})\Theta( x_t^{(1)}-x_t)}{ L_1}\frac{r}{R},\nonumber\\
\bar{A}_0&=&0.
\eeq

Thus, for configurations with zero net topological charge containing pairs of instanton-anti-instanton separated in time, I find the interior gauge configuration to be perfectly smooth. The same goes for more spatially localized temporally separated instanton-anti-instanton pairs.

\noindent
{\bf Four dimensional analog in $SU(N)$ gauge fields:}
Here I will begin with one of the simplest integer topological charge configurations in Euclidean $3+1$ dimensions (i.e. $n=2$) for $SU(2)$ gauge fields. I will denote the spatial directions as $x_1, x_2, x_3$ and Euclidean time as $x_t$. To match onto our conventions for the five dimensional disk construction discussed earlier, we can identify $x_1$ with $w$ and $x_2$ with $z$. Away from the temporal boundary, we can consider the following ansatz:
\beq
A_i=\frac{-ig(x_t)}{g_{\text{YM}}} \,\,\Omega^{-1}\partial_i\Omega ,\,\, A_0=0
\label{gsu2}
\eeq
where $i=1,2,3$,
\beq
g(x_t)=\Theta(x_t-x_t^{(0)})
\eeq
 and
\beq
\Omega=\cos(f(\rho))I_{2\times 2}-\sum_{i=1,2,3} i\sin(f(\rho))\sigma_i \frac{x_i}{\rho}.
\label{ogsu2}
\eeq
Here $\rho=\sqrt{\sum_{i=1}^3 x_i^2}$ and $\sigma_i$ are Pauli matrices. 
The function $f(\rho)$ behaves in the following way
\beq
f(\rho)=\begin{cases}
-N\pi \,\,\,\,\,\text{for}\,\, \rho\ll\rho_0\\
0\,\,\,\,\,\,\,\,\,\,\,\,\,\,\,\,\text{for}\,\, \rho\gg\rho_0
\end{cases}
\eeq
where $N$ is an integer which will coincide with the topological charge later in the calculation. We can also specify an exact functional form for $f(\rho)$, e.g. 
\beq
f(\rho)=- N\pi \left(1-S\left(\frac{\rho-\rho_0}{\rho_1-\rho_0}\right)\right),\nonumber\\
S(s)=\frac{\eta(s)}{\eta(s)+\eta(1-s)},\nonumber\\\nonumber\\
\eta(s)=\begin{cases}
0\,\,\,\,\,\,\,\,\,\,\,\,\,\,\,\text{for}\,\,\,\, s\leq 0\\
e^{-1/s}\,\,\,\,\text{for}\,\,\,\, s> 0.
\end{cases}
\eeq
 The gauge config in Eq. \ref{gsu2} is pure gauge for $x_t\ll x_t^{(0)}$ and $x_t\gg x_t^{(0)}$. The instanton number for the gauge field configuration is given by 
\beq
\mathcal{N}=\frac{g_{\text{YM}}^2}{64\pi^2}\int d^4x\,\, \epsilon_{\mu\nu\sigma\lambda}F^a_{\mu\nu}F^a_{\sigma\lambda}.
\label{intref}
\eeq
We can compute the above integral away from the temporal boundary, e.g. between $x_t=x_t^{(1)}$ and $x_t^{(2)}$ where $x_t^{(1)}\ll x_t^{(0)}$ and $x_t^{(2)}\gg x_t^{(0)}$ while also satisfying $x_t^{(1)}>0,   x_t^{(2)}<L_t$. Note that the ansatz for $x_t^{(1)}$ and $x_t^{(2)}$ approach pure gauge fields with an integral over all space between $x_t^{(1)}$ and $x_t^{(2)}$, $\mathcal{N}$ can be written as
\beq
\mathcal{N}&=&
 w(x_t^{(2)})- w(x_t^{(1)})=N
\label{nint}
\eeq
where 
\beq
w(x_t)=\frac{\left(ig_{\text{YM}}\right)^3}{24\pi^2}\int d^3x \,\,\epsilon_{ijk}\text{Tr}(A_i(x_t) A_j(x_t) A_k(x_t)). \nonumber\\
\eeq
It is clear that $w(x_t^{(1, 2)})$ are winding numbers of the gauge fields at time $x_t^{(1, 2)}$. Hence, they are integers. Consequently, $\mathcal{N}$ is expected to be an integer. And in fact I find it to be so and it equals $N$. From this point, we can focus on $N=\pm 1$ configurations and combinations thereof. 
I can now focus on the temporal boundary. It is clear from the ansatz in Eq. \ref{gsu2} that $A_\mu$ is not periodic in $x_t$ if we extend the ansatz all the way to the temporal boundary. However, they are gauge equivalent and this can be seen with

\beq
A_i(L_t, x_i)&=&A_i(0, x_i)-\frac{i}{g_{\text{YM}}}\Omega^{-1}(L_t, x)\partial_i \Omega(L_t,x),\nonumber\\
A_0(x_t, x_i)&=&\frac{i}{g_{\text{YM}}}(\log\Omega(L_t, x_i))\tilde{\delta}(x_t).
\label{last}
\eeq
Again, the above can be translated to the lattice. As before, $L_t$ on the lattice has to be an integer and the gauge links can then be expressed as 
\beq
U_i(x_t, x)=e^{g(x_t)\log(\Omega^{-1}(x)\Omega(x+\hat{i}))}
\eeq
for $x_t=0, \cdots L_{t}-1$
and 
\beq
 U_0(x_t, x_1)=\begin {cases}1\hspace{.9 in} \text{for $x_t=0, \cdots L_t-2$}\\
 \Omega^{-1}(L_t-1, x_i)\hspace{0.4 in} \text{for $x_t=L_t-1$}.
 \end{cases}
\eeq
where I can simply use the continuum ansatz for $A_i$ and $\Omega$ from Eq. \ref{gsu2} and \ref{ogsu2}.
One can now extend the domain of the integral in Eq. \ref{intref} to all of $x_t$ including the temporal boundary and the result equals the RHS of Eq. \ref{nint}. This is the instanton number for this configuration.

We can now assess the fate of such instanton configurations when they are realized on the boundary of our $4+1$ dimensional manifold with the identification $x_3\rightarrow R\phi$. The $1+1$ dimensional discussion and the absence of any topological obstruction to extending smooth boundary gauge fields into the interior suggests that we don't expect any singularity in the interior away from the temporal boundary. An explicit demonstration of this is found in the regime where Abelianization holds, where we can drop the nonlinear terms. In that case, Eq.  \ref{ar}, \ref{aphi}, \ref{az} and \ref{aw} with \ref{eqda}, produce non-singular solutions as in Eq. \ref{asol} and \ref{asol2}. However, the Abelianization of the equations breaks down when $g(x_t)\sim 1$, near $r=R$, i.e. for time slices for which the boundary data is pure gauge near the boundary. So, even though we don't expect singularities for these $x_t$, an explicit demonstration for all boundary values will require numerical work. 
At the temporal boundary, $U_0$ remains periodic in $x_3$, i.e. single-valued in $\phi$ for $r=R$. However, as the Abelian calculation suggests, we expect there to be a singularity in the interior at the temporal boundary. Thus, the gauge fields do not remain smooth for $r<R$ at the temporal boundary.

Let's consider temporally separated instanton-anti-instanton pair. One of the simplest representative ansatze  that one can write is to use the expression for gauge field in Eq. \ref{gsu2} for all $x_t$ and replace the function $g(x_t)$ with a box function
with support over some $x_t^{(0)}$ to $x_t^{(1)}$
given by
\beq
g(x_t)\rightarrow\Theta(x_t-x_t^{(0)})\Theta(x_t^{(1)}-x_t)
\eeq
with $L_t>x_{t}^{(1)}>x_{t}^{(0)}>0$.
For such a configuration, the gauge field is continued into the bulk to all $r$ without any singularity at any $r$ for any $x_t$ while retaining smooth gauge links at the temporal boundary.  

We can now consider spatially separated instanton-anti-instanton pair configuration. The simplest ansatz we can construct is
\beq
\Omega=\Omega_1\Omega_2,\,\, 
A_i=-\frac{i}{g_{\text{YM}}}\Theta(x_t-x_t^{(0)})\Omega^{-1}\partial_i\Omega, A_0=0\nonumber\\
\label{su23}
\eeq
away from the temporal boundary 
with 
\beq
\Omega_1&=&\cos(f(\rho))I_{2\times 2}-\sum_{i=1,2,3} i\sin(f(\rho))\sigma_i \frac{x_i}{\rho}\nonumber\\
\Omega_2&=&\cos(\tilde{f}(\tilde{\rho}))I_{2\times 2}-\sum_{i=1,2,3} i\sin(\tilde{f}(\tilde{\rho}))\sigma_i \frac{x_i-\tilde{x}_i}{\tilde{\rho}}\nonumber\\
\rho &=&\sqrt{\sum_{i=1,2,3}x_i^2},\,\, \tilde{\rho}=\sqrt{\sum_{i=1,2,3}(x_i-\tilde{x}_i)^2}
\eeq
where 
\beq
\tilde{f}(\tilde\rho)=\pi \left(1-S\left(\frac{\tilde\rho-\rho_0}{\rho_1-\rho_0}\right)\right).\nonumber\\
\eeq
$x_i=0, x_t=x_t^{(0)}$ and $x_i=\tilde{x}_i, x_t=x_t^{(0)}$ are the centers of the  two configurations.

To extend the above gauge fields to the temporal boundary, I can deform
\beq
\Omega_1&=&\cos(f(\rho^{(0)}))I_{2\times 2}\nonumber\\
&-&\sum_{i=1,2,3} i\sin(f(\rho^{(0)}))\sigma_i \frac{x_i-h(x_t)x^{(0)}_i}{\rho^{(0)}}\nonumber\\\nonumber\\
\Omega_2&=&\cos(\tilde{f}(\tilde{\rho}^{(0)}))I_{2\times 2}\nonumber\\
&-&\sum_{i=1,2,3} i\sin(\tilde{f}(\tilde{\rho}^{(0)}))\sigma_i \frac{x_i-\tilde{x}_i(1-h(x_t))-x_i^{(0)}h(x_t)}{\tilde{\rho}^{(0)}}\nonumber\\
\eeq
where $h$ changes continuously from $0$ to $1$ around $x_t\approx l+ x_t^{0}$, over a width of $\alpha^{-1}$ where $L_t- x_t^{(0)}\gg l \gg \alpha^{-1}> 0$  e.g. I could pick $h(x_t)$ to be, 
\beq
h(x_t)\rightarrow \frac{1+\tanh\left(\alpha\left(x_t-x_t^{(0)}-l\right)\right)}{2}.
\eeq
 Also,
$\rho^{(0)}$ and $\tilde{\rho}^{(0)}$ are defined as
\beq
\rho^{(0)}&=&\sqrt{\sum_i\left(x_i-h(x_t)x^{(0)}_i\right)^2}\nonumber\\
\tilde{\rho}^{(0)}&=&\sqrt{\sum_i\left(x_i-\tilde{x}_i(1-h(x_t))-x_i^{(0)}h(x_t)\right)^2}.
\eeq

Thus, near  $x_t=x_t^{(0)}+l\gg x_t^{(0)}$, where spatial field strengths are approximately zero, we bring the centers of $\Omega_1$ and $\Omega_2$ close together, so they coincide. For $L_t\geq x_t>x_t^{(0)}+l$,  $\Omega$, which equals $\Omega_1\Omega_2$, approaches the identity. This configuration, when placed on the $r=R$ boundary of the $n=2$, $2n+1$  dimensional manifold has no discontinuity anywhere. 
In addition, when the gauge field equations Eq.\ref{eqmod} are solved for the interior $r<R$, we expect non-singular solutions for all  $x_t$ at any $w, z, \phi$ and any $r$. The same holds for instanton-anti-instanton pairs separated in space and time.

Note that, while we have chosen to work with a certain form of the gauge field ansatz, similar arguments can be extended to other types of instanton configurations in various gauge choices. E.g. the above analysis goes through the same way for BPST instantons.\\

\noindent
{\bf Implications for QCD subsector, $\eta'$ physics and the $\theta$ angle:} As the discussion above shows, we expect the prescription in \ref{eqmod}, with the gauge choice of Eq. \ref{eqda} to exhibit a very useful feature: it produces 
perfectly smooth gauge field configuration in the interior of the disk for all of $x_t$ and $r, \phi, w, z$ when the boundary gauge configuration does not have any net topological charge, even if it contains individual instanton-anti-instanton pairs. In addition, when the net boundary instanton number is nonzero, the interior will contain singularity. But the singularity is pushed to the temporal boundary. Therefore, just as in the case of the EOM flow, for a boundary gauge configuration with $n_{\text{i}}$ instantons and $\bar{n}_{\text{i}}$ anti-instantons, we will get $n_{\text{i}}+\bar{n}_{\text{i}}$ approximate boundary ($r=R$) chiral zero modes and $|n_{\text{i}}-\bar{n}_{\text{i}}|$ interior ($r<R$) mirror zero modes at the temporal boundary. As a result, the instanton contribution to the fermion partition function is analogous to \eq{isum}, in particular the $n_{\text{i}}=\bar{n}_{\text{i}}$ contribution, i.e. $Z_3$ is analogous. Thus, all of the analysis in \cite{Kaplan:2024ezz} resulting in the $\eta'$ EFT of \eq{Leta} applies to this case as well. In other words,  \eq{Leta} describes the $\eta'$ EFT for the flow prescription in Eq. \ref{eqmod} and \ref{eqda}. Thus, we expect that Eq. \ref{eqmod} and \ref{eqda} will lead to an $\eta'$ that is not a GB, just as in the case of the $2n+1$ dimensional EOM lift. All of the finite-momentum modes of $\eta'$ are gapped whereas the zero momentum mode remains gapless. The effect of the latter would be to turn the $\theta$ angle unphysical following the conclusion of \cite{Kaplan:2024ezz}.

\section{conclusion}
The disk construction of chiral gauge theories in \cite{Kaplan:2023pvd, Kaplan:2024ezz, Kaplan:2023pxd} provided the framework for a Euclidean formulation for CGT. In fact, its predictions for the QCD subsector of the SM can be tested in a Euclidean lattice simulation. However, the full CGT, including the chiral sector cannot be simulated on a Euclidean lattice due to the sign problem. This provides one of the central motivations of this paper. In addition, while the predictions of \cite{Kaplan:2024ezz} for the QCD subsector can be tested on a Euclidean lattice, an explicit Hamiltonian will clarify its connection with real time physics. Such a connection is highly desirable due to one of the predictions of \cite{Kaplan:2024ezz}, that the disk construction does not suffer from the strong CP problem.

Among the possible directions for future work, the following stand out. This paper has proposed a prescription to eliminate 
a central obstruction to constructing an explicit Hamiltonian for the disk, thereby opening the possibility of exploring generic CGTs with a Euclidean sign problem using a quantum simulation. Thus,
it will be useful to construct the full explicit Hamiltonian of the disk both for chiral gauge theories and vectorlike theories using this prescription. It will be interesting to test lower-dimensional CGTs within this prescription to test its efficacy.
In addition, the proposal was shown to admit smooth and non-singular bulk gauge fields within the weak coupling regime for the non-Abelian theory and exactly for Abelian CGT. Exploring these conclusions in a non-perturbative setting through numerics is worthwhile. Similarly, it may be worthwhile exploring other possible gauge choices to explore the smoothness of interior solutions for those other choices. 
 A second direction is to entertain the possibility of elevating the disk theory from a tool engineered for a CGT formulation to potentially a realistic description of the universe. E.g. it may be interesting to 
explore whether the disk boundary can be compatible with the phenomenology and cosmology of the physical universe. The framework presented in this paper can facilitate the exploration of such questions. Finally, it is worthwhile exploring if there are other pathways to a  Hamiltonian formulation connecting to the disk framework and whether they result in  phenomenology that is distinct from \cite{Kaplan:2023pxd} and this paper. 
\section*{Acknowledgments}
I would like to thank Jinlong Dang, David Kaplan, Ryan Plestid, Hersh Singh and Marina Marinkovic for insightful discussions and acknowledge support from the U.S.~Department of Energy, grant number DE-SC0026815. I am also grateful to CERN for hosting me during the summer of 2026 during which some of these ideas were conceived. ChatGPT (GPT-5.6) was used to assist in checking the formulas appearing in the paper, ChatGPT (GPT-6.1) to spot typos which the I corrected and find some references which I verified. 
\begin{widetext}
\appendix
\section{}
\label{app1}
\subsection{Solution for $\bar{A}_r, \bar{A}_\phi, \bar{A}_w, \bar{A}_z$}
To solve Eq. \ref{ar}, \ref{aphi}, \ref{aw}, \ref{az} and \ref{eqda} I consider the Fourier mode with $(m, p_w, p_z)$ at time $x_t$. If I denote 
\beq
\partial_i \bar{A}_i=g(r) e^{i(m\phi+ p_z z+ p_w w)}
\eeq
I can write 
\beq
&&\Delta \left(g(r)e^{i(m\phi+ p_z z+ p_w w)}\right)=0\nonumber\\
&\implies&\frac{1}{r}\partial_r \left(r \frac{\partial g}{\partial r}\right)e^{i(m\phi+ p_z z+ p_w w)}+g(r)\left(\frac{\partial_\phi^2}{r^2}+\partial_z^2+\partial_w^2\right)e^{i(m\phi+ p_z z+ p_w w)}=0
\nonumber\\
&\implies & \frac{1}{r}\partial_r \left(r \frac{\partial g}{\partial r}\right)-\left(\frac{m^2}{r^2}+p_z^2+p_w^2\right)g(r)=0.
\eeq
and this is solved by
\beq
g(r)=  C_3' I[m, p\, r]+\tilde{C}_3 K[m, p \,r]
\eeq
where $I$ and $K$ are modified Bessel functions of the first and the second kind and  $p=\sqrt{p_z^2+p_w^2}$.
To maintain non-singular solutions, I choose 
$\tilde{C}_3=0$ such that 
\beq
g(r)=  C_3' I[m, p\, r]
\eeq
which lets me write
\beq
\partial_i A_i=C_3' I[m, p\, r]e^{i(m\phi+p_z z+p_w w)}.
\eeq
I will now define $C_3'=2C_3 p$ for future convenience.\\

Now I consider the  $\hat{z}$ component equation to make it more compact:
\beq
&&\frac{1}{r}\partial_r\left(r \partial_z \bar{A}_r\right)-\frac{1}{r}\partial_r\left(r \partial_r \bar{A}_z\right)+\frac{1}{r}\partial_\phi\partial_z \bar{A}_\phi -\frac{1}{r^2}\partial_\phi^2 \bar{A}_z+\partial_z\partial_w \bar{A}_w-\partial_w^2 \bar{A}_z=0\nonumber\\
&\implies&\partial_z\left(\frac{1}{r}\partial_r\left(r  \bar{A}_r\right)+\frac{1}{r}\partial_\phi \bar{A}_\phi+\partial_w \bar{A}_w+\partial_z \bar{A}_z\right)-\left(\frac{1}{r}\partial_r(r \partial_r \bar{A}_z)+\frac{1}{r^2}\partial_\phi^2 \bar{A}_z+\partial_z^2 \bar{A}_z+\partial_w^2\bar{A}_z\right)=0\nonumber\\
&\implies& \partial_z (\partial_i \bar{A}_i)-\Delta \bar{A}_z=0 \implies\Delta \bar{A}_z =\partial_z (2C_3\, p\, I[m, p\, r]e^{i(m\phi+p_z z+p_w w)}).
\label{az0}
\eeq

Demanding non-singular solutions at the origin, Eq. \ref{az0} is solved by
\beq
\bar{A}_z=\left((i p_z C_1+\tilde{C}_1^{(T)}) I[m, p\, r]+ i p_z r \,C_3 I[m+1, p\, r]\right)e^{i(m\phi+p_z z+p_w w)}.
\label{az1}
\eeq

Similarly I can write the general and non-singular solution for $\bar{A}_w$ as

\beq
\bar{A}_w=\left((i p_w C_2+\tilde{C}_2^{(T)}) I[m, p\, r]+ i p_w r \,C_3 I[m+1, p\, r]\right)e^{i(m\phi+p_z z+p_w w)}
\label{aw1}. 
\eeq

I will come back to these equations later.

I now consider the $\hat{r}$ component and $\hat{\phi}$ component equations.
To simplify them, the following manipulation is useful:
\beq
\frac{\partial}{\partial r}\left(\frac{1}{r}\frac{\partial \bar{A}_r}{\partial\phi}\right)=-\frac{2}{r^2}\frac{\partial \bar{A}_r}{\partial\phi}+\frac{1}{r}\partial_\phi\left(\frac{1}{r}\partial_r(r \bar{A}_r)\right).
\eeq

The $\hat{\phi}$ equation, Eq. \ref{aphi} becomes:
\beq
&&\frac{1}{r}\partial_\phi\left(\frac{1}{r}\partial_r\left(r  \bar{A}_r\right)+\frac{1}{r}\partial_\phi \bar{A}_\phi+\partial_w \bar{A}_w+\partial_z \bar{A}_z\right)-\left(\partial_z^2 \bar{A}_\phi+\partial_w^2 \bar{A}_\phi+\partial_r\left(\frac{1}{r}\partial_r(r \bar{A}_\phi)\right)+\frac{1}{r^2}\partial_\phi^2 \bar{A}_\phi+\frac{2}{r^2}\partial_\phi \bar{A}_r\right)=0\nonumber\\
&\implies &\left(\partial_z^2 \bar{A}_\phi+\partial_w^2 \bar{A}_\phi+\partial_r\left(\frac{1}{r}\partial_r(r \bar{A}_\phi)\right)+\frac{1}{r^2}\partial_\phi^2 \bar{A}_\phi\right)+\frac{2}{r^2}\partial_\phi \bar{A}_r=\frac{1}{r}\partial_\phi (\partial_i \bar{A}_i)\nonumber\\
&\implies & \left(\partial_z^2 \bar{A}_\phi+\partial_w^2 \bar{A}_\phi+\partial_r\left(\frac{1}{r}\partial_r(r \bar{A}_\phi)\right)+\frac{1}{r^2}\partial_\phi^2 \bar{A}_\phi\right)+\frac{2}{r^2}\partial_\phi \bar{A}_r=\frac{im}{r}(2 C_3 p)I[m, p\, r]e^{i(m\phi+p_z z+p_w w)}
\eeq
Similarly, the $\hat{r}$ equation, Eq. \ref{ar} becomes:
\beq
&&\frac{1}{r}\partial_\phi\left(\frac{\bar{A}_\phi}{r}+\partial_r \bar{A}_\phi\right)-\left(\frac{1}{r^2}\partial_\phi^2+\partial_z^2+\partial_w^2\right)\bar{A}_r+\partial_r(\partial_z \bar{A}_z+\partial_w \bar{A}_w)=0\nonumber\\
&\implies & \frac{1}{r^2}\partial_\phi \bar{A}_\phi+\partial_r\left(\frac{1}{r}\partial_\phi \bar{A}_\phi\right)+\frac{1}{r^2}\partial_\phi \bar{A}_\phi-\left(\frac{1}{r^2}\partial_\phi^2+\partial_z^2+\partial_w^2\right)\bar{A}_r+\partial_r(\partial_z \bar{A}_z+\partial_w \bar{A}_w)=0\nonumber\\
&\implies & \frac{2}{r^2}\partial_\phi \bar{A}_\phi+\partial_r\left(\frac{1}{r}\partial_r\left(r  \bar{A}_r\right)+\frac{1}{r}\partial_\phi \bar{A}_\phi+\partial_w \bar{A}_w+\partial_z \bar{A}_z\right)-\left(\frac{1}{r^2}\partial_\phi^2 \bar{A}_r+\partial_z^2 \bar{A}_r+\partial_w^2 \bar{A}_r+\partial_r\left(\frac{1}{r}\partial_r(r \bar{A}_r)\right)\right)=0\nonumber\\
&\implies & \left(\partial_z^2 \bar{A}_r+\partial_w^2 \bar{A}_r+\frac{1}{r^2}\partial_\phi^2 \bar{A}_r-\frac{\bar{A}_r}{r^2}+\partial_r^2 \bar{A}_r+\frac{1}{r}\partial_r \bar{A}_r\right)-\frac{2}{r^2}\partial_\phi \bar{A}_\phi
=\partial_r(\partial_i \bar{A}_i)\nonumber\\
&&\hspace{3.75 in}=C_3p^2 \left(I[m-1,p r]+I[m+1, p r]\right)e^{i(m\phi+p_z z+p_w w)}.\nonumber\\\eeq

For convenience I write
\beq
\bar{A}_r=u e^{i(m\phi+p_z z+p_w w)}\nonumber\\
\bar{A}_\phi=v e^{i(m\phi+p_z z+p_w w)}
\label{uv}
\eeq
which turns the $\hat{r}$ and $\hat{\phi}$ equations into:
\beq
\left(\partial_r^2+\frac{1}{r}\partial_r-p^2-\frac{(m^2+1)}{r^2}\right)u-\frac{2im}{r^2}v&=&C_3\,p^2 \left(I[m-1,p r]+I[m+1, p r]\right)\nonumber\\
\left(\partial_r^2+\frac{1}{r}\partial_r-p^2-\frac{(m^2+1)}{r^2}\right)v+\frac{2im}{r^2}u&=&\frac{im}{r}(2C_3\,p )I[m,p r].
\label{uvrr}\nonumber\\
\eeq
 Further defining 
\beq
u+i v= \bar{A}_+,\,\,\,\, u-i v= \bar{A}_-,
\eeq
I solve Eq. \ref{uvrr} and demanding non-singular solutions, I obtain

\beq
\bar{A}_+= C_+ I[m+1, p\, r]+ C_3 r \frac{d}{dr}(I[m+1, p \, r])\nonumber\\
\bar{A}_-=C_- I[m-1, p\, r]+ C_3 r \frac{d}{dr}(I[m-1, p\, r])
\eeq

which leads to 
\beq
\bar{A}_r=\left(\frac{C_+ I[m+1, p\, r]+C_- I[m-1, p\, r]+ C_3 r \left(\frac{d}{dr}(I[m+1, p \, r])+ \frac{d}{dr}(I[m-1, p \, r])\right)}{2}\right)e^{i(m\phi+p_w w+p_z z)}\nonumber\\
\bar{A}_\phi=\left(\frac{C_+ I[m+1, p\, r]-C_- I[m-1, p\, r]+ C_3 r \left(\frac{d}{dr}(I[m+1, p \, r])- \frac{d}{dr}(I[m-1, p \, r])\right)}{2i}\right)e^{i(m\phi+p_w w+p_z z)}.
\label{afin1}
\eeq

Similarly defining $(i p_z C_1+\tilde{C}_1^{(T)})=\tilde{C}_1$ and $(i p_w C_2+\tilde{C}_2^{(T)})=\tilde{C}_2$ I can rewrite the solutions in Eq. \ref{az1} and \ref{aw1} as
\beq
\bar{A}_z=\left(\tilde{C}_1 I[m, p\, r]+ i p_z r \,C_3 I[m+1, p\, r]\right)e^{i(m\phi+p_z z+p_w w)}\nonumber\\
\bar{A}_w=\left(\tilde{C}_2 I[m, p\, r]+ i p_w r \,C_3 I[m+1, p\, r]\right)e^{i(m\phi+p_z z+p_w w)}.
\label{afin2}
\eeq
Eq. \ref{afin1} and \ref{afin2} are the solutions found in the main text in Eq. \ref{asol} and \ref{asol2}. 

\subsection{Green's function for $\bar{A}_0$}
I now solve for the Green's function of Eq. \ref{G} to obtain the $\bar{A}_0$ functional. 
For both $r>\rho$ and $r<\rho$ the RHS of Eq. \ref{G} is zero.\\ 

\noindent
For $r>\rho$ 
we write the Green's function as 
\beq
G(r,\rho)=D_{1,>}(\rho) I[m, p\, r]+D_{2,>}(\rho) K[m, p\, r]
\eeq\\
 and for $r<\rho$ 

\beq
G(r,\rho)=D_{1,<}(\rho) I[m, p\, r]+D_{2,<}(\rho) K[m, p\, r].
\eeq\\

For $r>\rho$, I impose $G=0$ for $r=R$,  leading to
\beq
D_{2,>}(\rho)=\frac{-D_{1,>}(\rho)I[m, p R]}{K[m, p R]}. 
\eeq

Similarly, for $r<\rho$, I demand $G$ be non-singular at $r=0$ which gives
\beq
D_{2,<}(\rho)=0.
\eeq

Thus for $r>\rho$, I have
\beq
G(r, \rho)=D_{1,>}(\rho)\left(I[m, p r]-\frac{I[m, p R]}{K[m, p R]}K[m, p r]\right)
\eeq
and for $r<\rho$
\beq
G(r, \rho)=D_{1,<}(\rho)I[m, p r].
\eeq

Demanding continuity at $r=\rho$ I get
\beq
D_{1,<}(\rho)I[m, p \rho]=D_{1,>}(\rho)\left(I[m, p\rho]-\frac{I[m, p R]}{K[m, p R]}K[m, p\rho]\right).
\eeq

This leads to the expression for the Green's function
\beq
G(r,\rho)=\begin{cases}
D_{1,>}(\rho)\frac{\left(I[m, p\rho]-\frac{I[m, pR]}{K[m, p R]}K[m, p\rho]\right)}{I[m, p\rho]}I[m, p r]\,\,\,\,\,\,\,\,\text{for}\,\, r<\rho\\\\
D_{1,>}(\rho)\left(I[m, pr]-\frac{I[m, pR]}{K[m, p R]}K[m, pr]\right)\,\,\,\,\,\,\,\,\text{for}\,\, r>\rho.
\end{cases}
\eeq

I then integrate both sides of Eq \ref{G}, from $r=\rho_-\equiv \rho-|\delta|$ to $r=\rho_+\equiv \rho+|\delta|$, with $\delta\rightarrow 0$, which gives 
\beq
D_{1,>}=\frac{I[m, p \rho]K[m, p R]}{I[m, p R]}. 
\eeq

From here, I get the final expression for the Green's function
\beq
G(r,\rho)=\begin{cases}
 -I[m, p r]\left(K[m, p\rho]-\frac{K[m, p R]}{I[m, p R]}I[m , p\rho]\right)\,\,\,\,\text{for}\,\,\,\,r<\rho\\
-I[m, p\rho]\left(K[m, p r]-\frac{K[m, p R]}{I[m, p R]}I[m, p r]\right)\,\,\,\,\text{for}\,\,\,\,r>\rho.
\end{cases}
\eeq

\end{widetext}

\bibliography{refs.bib}

\begin{thebibliography}{45}%
\makeatletter
\providecommand \@ifxundefined [1]{%
 \@ifx{#1\undefined}
}%
\providecommand \@ifnum [1]{%
 \ifnum #1\expandafter \@firstoftwo
 \else \expandafter \@secondoftwo
 \fi
}%
\providecommand \@ifx [1]{%
 \ifx #1\expandafter \@firstoftwo
 \else \expandafter \@secondoftwo
 \fi
}%
\providecommand \natexlab [1]{#1}%
\providecommand \enquote  [1]{``#1''}%
\providecommand \bibnamefont  [1]{#1}%
\providecommand \bibfnamefont [1]{#1}%
\providecommand \citenamefont [1]{#1}%
\providecommand \href@noop [0]{\@secondoftwo}%
\providecommand \href [0]{\begingroup \@sanitize@url \@href}%
\providecommand \@href[1]{\@@startlink{#1}\@@href}%
\providecommand \@@href[1]{\endgroup#1\@@endlink}%
\providecommand \@sanitize@url [0]{\catcode `\\12\catcode `\$12\catcode
  `\&12\catcode `\#12\catcode `\^12\catcode `\_12\catcode `\%12\relax}%
\providecommand \@@startlink[1]{}%
\providecommand \@@endlink[0]{}%
\providecommand \url  [0]{\begingroup\@sanitize@url \@url }%
\providecommand \@url [1]{\endgroup\@href {#1}{\urlprefix }}%
\providecommand \urlprefix  [0]{URL }%
\providecommand \Eprint [0]{\href }%
\providecommand \doibase [0]{https://doi.org/}%
\providecommand \selectlanguage [0]{\@gobble}%
\providecommand \bibinfo  [0]{\@secondoftwo}%
\providecommand \bibfield  [0]{\@secondoftwo}%
\providecommand \translation [1]{[#1]}%
\providecommand \BibitemOpen [0]{}%
\providecommand \bibitemStop [0]{}%
\providecommand \bibitemNoStop [0]{.\EOS\space}%
\providecommand \EOS [0]{\spacefactor3000\relax}%
\providecommand \BibitemShut  [1]{\csname bibitem#1\endcsname}%
\let\auto@bib@innerbib\@empty
\bibitem [{\citenamefont {Luscher}(1999)}]{Luscher:1998du}%
  \BibitemOpen
  \bibfield  {author} {\bibinfo {author} {\bibfnamefont {M.}~\bibnamefont
  {Luscher}},\ }\bibfield  {title} {\bibinfo {title} {{Abelian chiral gauge
  theories on the lattice with exact gauge invariance}},\ }\href
  {https://doi.org/10.1016/S0550-3213(99)00115-7} {\bibfield  {journal}
  {\bibinfo  {journal} {Nucl. Phys. B}\ }\textbf {\bibinfo {volume} {549}},\
  \bibinfo {pages} {295} (\bibinfo {year} {1999})},\ \Eprint
  {https://arxiv.org/abs/hep-lat/9811032} {arXiv:hep-lat/9811032} \BibitemShut
  {NoStop}%
\bibitem [{\citenamefont {Kadoh}\ and\ \citenamefont
  {Kikukawa}(2008)}]{Kadoh:2007wz}%
  \BibitemOpen
  \bibfield  {author} {\bibinfo {author} {\bibfnamefont {D.}~\bibnamefont
  {Kadoh}}\ and\ \bibinfo {author} {\bibfnamefont {Y.}~\bibnamefont
  {Kikukawa}},\ }\bibfield  {title} {\bibinfo {title} {{A Simple construction
  of fermion measure term in U(1) chiral lattice gauge theories with exact
  gauge invariance}},\ }\href {https://doi.org/10.1088/1126-6708/2008/02/063}
  {\bibfield  {journal} {\bibinfo  {journal} {JHEP}\ }\textbf {\bibinfo
  {volume} {02}},\ \bibinfo {pages} {063}},\ \Eprint
  {https://arxiv.org/abs/0709.3656} {arXiv:0709.3656 [hep-lat]} \BibitemShut
  {NoStop}%
\bibitem [{\citenamefont {Kikukawa}(2019)}]{Kikukawa:2017ngf}%
  \BibitemOpen
  \bibfield  {author} {\bibinfo {author} {\bibfnamefont {Y.}~\bibnamefont
  {Kikukawa}},\ }\bibfield  {title} {\bibinfo {title} {{On the gauge invariant
  path-integral measure for the overlap Weyl fermions in $\underline{16}$ of
  SO(10)}},\ }\href {https://doi.org/10.1093/ptep/ptz115} {\bibfield  {journal}
  {\bibinfo  {journal} {PTEP}\ }\textbf {\bibinfo {volume} {2019}},\ \bibinfo
  {pages} {113B03} (\bibinfo {year} {2019})},\ \Eprint
  {https://arxiv.org/abs/1710.11618} {arXiv:1710.11618 [hep-lat]} \BibitemShut
  {NoStop}%
\bibitem [{\citenamefont {Bock}\ \emph {et~al.}(1998)\citenamefont {Bock},
  \citenamefont {Golterman},\ and\ \citenamefont {Shamir}}]{Bock:1997ks}%
  \BibitemOpen
  \bibfield  {author} {\bibinfo {author} {\bibfnamefont {W.}~\bibnamefont
  {Bock}}, \bibinfo {author} {\bibfnamefont {M.~F.~L.}\ \bibnamefont
  {Golterman}},\ and\ \bibinfo {author} {\bibfnamefont {Y.}~\bibnamefont
  {Shamir}},\ }\bibfield  {title} {\bibinfo {title} {{Chiral fermions on the
  lattice through gauge fixing: Perturbation theory}},\ }\href
  {https://doi.org/10.1103/PhysRevD.58.034501} {\bibfield  {journal} {\bibinfo
  {journal} {Phys. Rev. D}\ }\textbf {\bibinfo {volume} {58}},\ \bibinfo
  {pages} {034501} (\bibinfo {year} {1998})},\ \Eprint
  {https://arxiv.org/abs/hep-lat/9801018} {arXiv:hep-lat/9801018} \BibitemShut
  {NoStop}%
\bibitem [{\citenamefont {Berkowitz}\ \emph {et~al.}(2024)\citenamefont
  {Berkowitz}, \citenamefont {Cherman},\ and\ \citenamefont
  {Jacobson}}]{Berkowitz:2023pnz}%
  \BibitemOpen
  \bibfield  {author} {\bibinfo {author} {\bibfnamefont {E.}~\bibnamefont
  {Berkowitz}}, \bibinfo {author} {\bibfnamefont {A.}~\bibnamefont {Cherman}},\
  and\ \bibinfo {author} {\bibfnamefont {T.}~\bibnamefont {Jacobson}},\
  }\bibfield  {title} {\bibinfo {title} {{Exact lattice chiral symmetry in 2D
  gauge theory}},\ }\href {https://doi.org/10.1103/PhysRevD.110.014510}
  {\bibfield  {journal} {\bibinfo  {journal} {Phys. Rev. D}\ }\textbf {\bibinfo
  {volume} {110}},\ \bibinfo {pages} {014510} (\bibinfo {year} {2024})},\
  \Eprint {https://arxiv.org/abs/2310.17539} {arXiv:2310.17539 [hep-lat]}
  \BibitemShut {NoStop}%
\bibitem [{\citenamefont {Baig}\ \emph {et~al.}(2026)\citenamefont {Baig},
  \citenamefont {Chen}, \citenamefont {Cherman},\ and\ \citenamefont
  {Neuzil}}]{Baig:2026zgb}%
  \BibitemOpen
  \bibfield  {author} {\bibinfo {author} {\bibfnamefont {S.~U.}\ \bibnamefont
  {Baig}}, \bibinfo {author} {\bibfnamefont {S.}~\bibnamefont {Chen}}, \bibinfo
  {author} {\bibfnamefont {A.}~\bibnamefont {Cherman}},\ and\ \bibinfo {author}
  {\bibfnamefont {M.}~\bibnamefont {Neuzil}},\ }\href@noop {} {\bibinfo {title}
  {{Bosonization versus the Nielsen-Ninomiya theorem}}} (\bibinfo {year}
  {2026}),\ \Eprint {https://arxiv.org/abs/2607.09935} {arXiv:2607.09935
  [hep-th]} \BibitemShut {NoStop}%
\bibitem [{\citenamefont {Lu}\ and\ \citenamefont {Shao}(2026)}]{Lu:2026itw}%
  \BibitemOpen
  \bibfield  {author} {\bibinfo {author} {\bibfnamefont {Z.}~\bibnamefont
  {Lu}}\ and\ \bibinfo {author} {\bibfnamefont {S.-H.}\ \bibnamefont {Shao}},\
  }\href@noop {} {\bibinfo {title} {{Fermionic Villain model with exact lattice
  chiral symmetries}}} (\bibinfo {year} {2026}),\ \Eprint
  {https://arxiv.org/abs/2608.02728} {arXiv:2608.02728 [hep-th]} \BibitemShut
  {NoStop}%
\bibitem [{\citenamefont {Seifnashri}(2026)}]{Seifnashri:2026ema}%
  \BibitemOpen
  \bibfield  {author} {\bibinfo {author} {\bibfnamefont {S.}~\bibnamefont
  {Seifnashri}},\ }\href@noop {} {\bibinfo {title} {{Exactly Solvable 1+1d
  Chiral Lattice Gauge Theories}}} (\bibinfo {year} {2026}),\ \Eprint
  {https://arxiv.org/abs/2601.14359} {arXiv:2601.14359 [hep-th]} \BibitemShut
  {NoStop}%
\bibitem [{\citenamefont {Thorngren}\ \emph {et~al.}(2026)\citenamefont
  {Thorngren}, \citenamefont {Preskill},\ and\ \citenamefont
  {Fidkowski}}]{Thorngren:2026ydw}%
  \BibitemOpen
  \bibfield  {author} {\bibinfo {author} {\bibfnamefont {R.}~\bibnamefont
  {Thorngren}}, \bibinfo {author} {\bibfnamefont {J.}~\bibnamefont
  {Preskill}},\ and\ \bibinfo {author} {\bibfnamefont {L.}~\bibnamefont
  {Fidkowski}},\ }\bibfield  {title} {\bibinfo {title} {{Chiral lattice gauge
  theories from symmetry disentanglers}},\ }\href
  {https://doi.org/10.1007/JHEP07(2026)271} {\bibfield  {journal} {\bibinfo
  {journal} {JHEP}\ }\textbf {\bibinfo {volume} {07}},\ \bibinfo {pages}
  {271}},\ \Eprint {https://arxiv.org/abs/2601.04304} {arXiv:2601.04304
  [hep-th]} \BibitemShut {NoStop}%
\bibitem [{\citenamefont {Wen}(2013)}]{Wen:2013ppa}%
  \BibitemOpen
  \bibfield  {author} {\bibinfo {author} {\bibfnamefont {X.-G.}\ \bibnamefont
  {Wen}},\ }\bibfield  {title} {\bibinfo {title} {{A lattice non-perturbative
  definition of an SO(10) chiral gauge theory and its induced standard
  model}},\ }\href {https://doi.org/10.1088/0256-307X/30/11/111101} {\bibfield
  {journal} {\bibinfo  {journal} {Chin. Phys. Lett.}\ }\textbf {\bibinfo
  {volume} {30}},\ \bibinfo {pages} {111101} (\bibinfo {year} {2013})},\
  \Eprint {https://arxiv.org/abs/1305.1045} {arXiv:1305.1045 [hep-lat]}
  \BibitemShut {NoStop}%
\bibitem [{\citenamefont {You}\ and\ \citenamefont {Xu}(2014)}]{You:2014sqa}%
  \BibitemOpen
  \bibfield  {author} {\bibinfo {author} {\bibfnamefont {Y.-Z.}\ \bibnamefont
  {You}}\ and\ \bibinfo {author} {\bibfnamefont {C.}~\bibnamefont {Xu}},\
  }\bibfield  {title} {\bibinfo {title} {{Symmetry Protected Topological States
  of Interacting Fermions and Bosons}},\ }\href
  {https://doi.org/10.1103/PhysRevB.90.245120} {\bibfield  {journal} {\bibinfo
  {journal} {Phys. Rev. B}\ }\textbf {\bibinfo {volume} {90}},\ \bibinfo
  {pages} {245120} (\bibinfo {year} {2014})},\ \Eprint
  {https://arxiv.org/abs/1409.0168} {arXiv:1409.0168 [cond-mat.str-el]}
  \BibitemShut {NoStop}%
\bibitem [{\citenamefont {You}\ and\ \citenamefont {Xu}(2015)}]{You:2014vea}%
  \BibitemOpen
  \bibfield  {author} {\bibinfo {author} {\bibfnamefont {Y.-Z.}\ \bibnamefont
  {You}}\ and\ \bibinfo {author} {\bibfnamefont {C.}~\bibnamefont {Xu}},\
  }\bibfield  {title} {\bibinfo {title} {{Interacting Topological Insulator and
  Emergent Grand Unified Theory}},\ }\href
  {https://doi.org/10.1103/PhysRevB.91.125147} {\bibfield  {journal} {\bibinfo
  {journal} {Phys. Rev. B}\ }\textbf {\bibinfo {volume} {91}},\ \bibinfo
  {pages} {125147} (\bibinfo {year} {2015})},\ \Eprint
  {https://arxiv.org/abs/1412.4784} {arXiv:1412.4784 [cond-mat.str-el]}
  \BibitemShut {NoStop}%
\bibitem [{\citenamefont {You}\ \emph {et~al.}(2014)\citenamefont {You},
  \citenamefont {BenTov},\ and\ \citenamefont {Xu}}]{You:2014oaa}%
  \BibitemOpen
  \bibfield  {author} {\bibinfo {author} {\bibfnamefont {Y.}~\bibnamefont
  {You}}, \bibinfo {author} {\bibfnamefont {Y.}~\bibnamefont {BenTov}},\ and\
  \bibinfo {author} {\bibfnamefont {C.}~\bibnamefont {Xu}},\ }\href@noop {}
  {\bibinfo {title} {{Interacting Topological Superconductors and possible
  Origin of $16n$ Chiral Fermions in the Standard Model}}} (\bibinfo {year}
  {2014}),\ \Eprint {https://arxiv.org/abs/1402.4151} {arXiv:1402.4151
  [cond-mat.str-el]} \BibitemShut {NoStop}%
\bibitem [{\citenamefont {Wang}\ and\ \citenamefont
  {Wen}(2018)}]{Wang:2018ugf}%
  \BibitemOpen
  \bibfield  {author} {\bibinfo {author} {\bibfnamefont {J.}~\bibnamefont
  {Wang}}\ and\ \bibinfo {author} {\bibfnamefont {X.-G.}\ \bibnamefont {Wen}},\
  }\bibfield  {title} {\bibinfo {title} {{A Solution to the 1+1D Gauged Chiral
  Fermion Problem}},\ }\href {https://doi.org/10.1103/PhysRevD.99.111501}
  {\bibfield  {journal} {\bibinfo  {journal} {Phys. Rev. D}\ }\textbf {\bibinfo
  {volume} {99}},\ \bibinfo {pages} {111501} (\bibinfo {year} {2018})},\
  \Eprint {https://arxiv.org/abs/1807.05998} {arXiv:1807.05998 [hep-lat]}
  \BibitemShut {NoStop}%
\bibitem [{\citenamefont {Catterall}(2021)}]{Catterall:2020fep}%
  \BibitemOpen
  \bibfield  {author} {\bibinfo {author} {\bibfnamefont {S.}~\bibnamefont
  {Catterall}},\ }\bibfield  {title} {\bibinfo {title} {{Chiral lattice
  fermions from staggered fields}},\ }\href
  {https://doi.org/10.1103/PhysRevD.104.014503} {\bibfield  {journal} {\bibinfo
   {journal} {Phys. Rev. D}\ }\textbf {\bibinfo {volume} {104}},\ \bibinfo
  {pages} {014503} (\bibinfo {year} {2021})},\ \Eprint
  {https://arxiv.org/abs/2010.02290} {arXiv:2010.02290 [hep-lat]} \BibitemShut
  {NoStop}%
\bibitem [{\citenamefont {Wang}\ and\ \citenamefont
  {You}(2022)}]{Wang:2022ucy}%
  \BibitemOpen
  \bibfield  {author} {\bibinfo {author} {\bibfnamefont {J.}~\bibnamefont
  {Wang}}\ and\ \bibinfo {author} {\bibfnamefont {Y.-Z.}\ \bibnamefont {You}},\
  }\bibfield  {title} {\bibinfo {title} {{Symmetric Mass Generation}},\ }\href
  {https://doi.org/10.3390/sym14071475} {\bibfield  {journal} {\bibinfo
  {journal} {Symmetry}\ }\textbf {\bibinfo {volume} {14}},\ \bibinfo {pages}
  {1475} (\bibinfo {year} {2022})},\ \Eprint {https://arxiv.org/abs/2204.14271}
  {arXiv:2204.14271 [cond-mat.str-el]} \BibitemShut {NoStop}%
\bibitem [{\citenamefont {Razamat}\ and\ \citenamefont
  {Tong}(2021)}]{Razamat:2020kyf}%
  \BibitemOpen
  \bibfield  {author} {\bibinfo {author} {\bibfnamefont {S.~S.}\ \bibnamefont
  {Razamat}}\ and\ \bibinfo {author} {\bibfnamefont {D.}~\bibnamefont {Tong}},\
  }\bibfield  {title} {\bibinfo {title} {{Gapped Chiral Fermions}},\ }\href
  {https://doi.org/10.1103/PhysRevX.11.011063} {\bibfield  {journal} {\bibinfo
  {journal} {Phys. Rev. X}\ }\textbf {\bibinfo {volume} {11}},\ \bibinfo
  {pages} {011063} (\bibinfo {year} {2021})},\ \Eprint
  {https://arxiv.org/abs/2009.05037} {arXiv:2009.05037 [hep-th]} \BibitemShut
  {NoStop}%
\bibitem [{\citenamefont {Nguyen}\ and\ \citenamefont
  {Singh}(2024)}]{Nguyen:2024wck}%
  \BibitemOpen
  \bibfield  {author} {\bibinfo {author} {\bibfnamefont {M.}~\bibnamefont
  {Nguyen}}\ and\ \bibinfo {author} {\bibfnamefont {H.}~\bibnamefont {Singh}},\
  }\href@noop {} {\bibinfo {title} {{Chiral symmetry and Atiyah-Patodi-Singer
  index theorem for staggered fermions}}} (\bibinfo {year} {2024}),\ \Eprint
  {https://arxiv.org/abs/2405.11348} {arXiv:2405.11348 [hep-lat]} \BibitemShut
  {NoStop}%
\bibitem [{\citenamefont {Ayyar}\ and\ \citenamefont
  {Chandrasekharan}(2016)}]{Ayyar:2015lrd}%
  \BibitemOpen
  \bibfield  {author} {\bibinfo {author} {\bibfnamefont {V.}~\bibnamefont
  {Ayyar}}\ and\ \bibinfo {author} {\bibfnamefont {S.}~\bibnamefont
  {Chandrasekharan}},\ }\bibfield  {title} {\bibinfo {title} {{Origin of
  fermion masses without spontaneous symmetry breaking}},\ }\href
  {https://doi.org/10.1103/PhysRevD.93.081701} {\bibfield  {journal} {\bibinfo
  {journal} {Phys. Rev. D}\ }\textbf {\bibinfo {volume} {93}},\ \bibinfo
  {pages} {081701} (\bibinfo {year} {2016})},\ \Eprint
  {https://arxiv.org/abs/1511.09071} {arXiv:1511.09071 [hep-lat]} \BibitemShut
  {NoStop}%
\bibitem [{\citenamefont {Ayyar}\ and\ \citenamefont
  {Chandrasekharan}(2015)}]{Ayyar:2014eua}%
  \BibitemOpen
  \bibfield  {author} {\bibinfo {author} {\bibfnamefont {V.}~\bibnamefont
  {Ayyar}}\ and\ \bibinfo {author} {\bibfnamefont {S.}~\bibnamefont
  {Chandrasekharan}},\ }\bibfield  {title} {\bibinfo {title} {{Massive fermions
  without fermion bilinear condensates}},\ }\href
  {https://doi.org/10.1103/PhysRevD.91.065035} {\bibfield  {journal} {\bibinfo
  {journal} {Phys. Rev. D}\ }\textbf {\bibinfo {volume} {91}},\ \bibinfo
  {pages} {065035} (\bibinfo {year} {2015})},\ \Eprint
  {https://arxiv.org/abs/1410.6474} {arXiv:1410.6474 [hep-lat]} \BibitemShut
  {NoStop}%
\bibitem [{\citenamefont {Catterall}(2024)}]{Catterall:2023nww}%
  \BibitemOpen
  \bibfield  {author} {\bibinfo {author} {\bibfnamefont {S.}~\bibnamefont
  {Catterall}},\ }\bibfield  {title} {\bibinfo {title} {{Lattice regularization
  of reduced K{\"a}hler-Dirac fermions and connections to chiral fermions}},\
  }\href {https://doi.org/10.21468/SciPostPhys.16.4.108} {\bibfield  {journal}
  {\bibinfo  {journal} {SciPost Phys.}\ }\textbf {\bibinfo {volume} {16}},\
  \bibinfo {pages} {108} (\bibinfo {year} {2024})},\ \Eprint
  {https://arxiv.org/abs/2311.02487} {arXiv:2311.02487 [hep-lat]} \BibitemShut
  {NoStop}%
\bibitem [{\citenamefont {Kaplan}(2024)}]{Kaplan:2023pxd}%
  \BibitemOpen
  \bibfield  {author} {\bibinfo {author} {\bibfnamefont {D.~B.}\ \bibnamefont
  {Kaplan}},\ }\bibfield  {title} {\bibinfo {title} {{Chiral Gauge Theory at
  the Boundary between Topological Phases}},\ }\href
  {https://doi.org/10.1103/PhysRevLett.132.141603} {\bibfield  {journal}
  {\bibinfo  {journal} {Phys. Rev. Lett.}\ }\textbf {\bibinfo {volume} {132}},\
  \bibinfo {pages} {141603} (\bibinfo {year} {2024})},\ \Eprint
  {https://arxiv.org/abs/2312.01494} {arXiv:2312.01494 [hep-lat]} \BibitemShut
  {NoStop}%
\bibitem [{\citenamefont {Kaplan}\ and\ \citenamefont
  {Sen}(2024)}]{Kaplan:2023pvd}%
  \BibitemOpen
  \bibfield  {author} {\bibinfo {author} {\bibfnamefont {D.~B.}\ \bibnamefont
  {Kaplan}}\ and\ \bibinfo {author} {\bibfnamefont {S.}~\bibnamefont {Sen}},\
  }\bibfield  {title} {\bibinfo {title} {{Weyl Fermions on a Finite Lattice}},\
  }\href {https://doi.org/10.1103/PhysRevLett.132.141604} {\bibfield  {journal}
  {\bibinfo  {journal} {Phys. Rev. Lett.}\ }\textbf {\bibinfo {volume} {132}},\
  \bibinfo {pages} {141604} (\bibinfo {year} {2024})},\ \Eprint
  {https://arxiv.org/abs/2312.04012} {arXiv:2312.04012 [hep-lat]} \BibitemShut
  {NoStop}%
\bibitem [{\citenamefont {Kaplan}(1992)}]{Kaplan:1992bt}%
  \BibitemOpen
  \bibfield  {author} {\bibinfo {author} {\bibfnamefont {D.~B.}\ \bibnamefont
  {Kaplan}},\ }\bibfield  {title} {\bibinfo {title} {{A Method for simulating
  chiral fermions on the lattice}},\ }\href
  {https://doi.org/10.1016/0370-2693(92)91112-M} {\bibfield  {journal}
  {\bibinfo  {journal} {Phys. Lett. B}\ }\textbf {\bibinfo {volume} {288}},\
  \bibinfo {pages} {342} (\bibinfo {year} {1992})},\ \Eprint
  {https://arxiv.org/abs/hep-lat/9206013} {arXiv:hep-lat/9206013} \BibitemShut
  {NoStop}%
\bibitem [{\citenamefont {Grabowska}\ and\ \citenamefont
  {Kaplan}(2016)}]{Grabowska:2015qpk}%
  \BibitemOpen
  \bibfield  {author} {\bibinfo {author} {\bibfnamefont {D.~M.}\ \bibnamefont
  {Grabowska}}\ and\ \bibinfo {author} {\bibfnamefont {D.~B.}\ \bibnamefont
  {Kaplan}},\ }\bibfield  {title} {\bibinfo {title} {{Nonperturbative Regulator
  for Chiral Gauge Theories?}},\ }\href
  {https://doi.org/10.1103/PhysRevLett.116.211602} {\bibfield  {journal}
  {\bibinfo  {journal} {Phys. Rev. Lett.}\ }\textbf {\bibinfo {volume} {116}},\
  \bibinfo {pages} {211602} (\bibinfo {year} {2016})},\ \Eprint
  {https://arxiv.org/abs/1511.03649} {arXiv:1511.03649 [hep-lat]} \BibitemShut
  {NoStop}%
\bibitem [{\citenamefont {Shamir}(1993)}]{Shamir:1993zy}%
  \BibitemOpen
  \bibfield  {author} {\bibinfo {author} {\bibfnamefont {Y.}~\bibnamefont
  {Shamir}},\ }\bibfield  {title} {\bibinfo {title} {{Chiral fermions from
  lattice boundaries}},\ }\href {https://doi.org/10.1016/0550-3213(93)90162-I}
  {\bibfield  {journal} {\bibinfo  {journal} {Nucl. Phys. B}\ }\textbf
  {\bibinfo {volume} {406}},\ \bibinfo {pages} {90} (\bibinfo {year} {1993})},\
  \Eprint {https://arxiv.org/abs/hep-lat/9303005} {arXiv:hep-lat/9303005}
  \BibitemShut {NoStop}%
\bibitem [{\citenamefont {Dang}\ \emph
  {et~al.}(2026{\natexlab{a}})\citenamefont {Dang}, \citenamefont {Karur},\
  and\ \citenamefont {Sen}}]{Dang:2026pkb}%
  \BibitemOpen
  \bibfield  {author} {\bibinfo {author} {\bibfnamefont {J.}~\bibnamefont
  {Dang}}, \bibinfo {author} {\bibfnamefont {R.}~\bibnamefont {Karur}},\ and\
  \bibinfo {author} {\bibfnamefont {S.}~\bibnamefont {Sen}},\ }\href@noop {}
  {\bibinfo {title} {{Gauge field flow for chiral gauge theories on a disk
  boundary}}} (\bibinfo {year} {2026}{\natexlab{a}}),\ \Eprint
  {https://arxiv.org/abs/2606.05305} {arXiv:2606.05305 [hep-lat]} \BibitemShut
  {NoStop}%
\bibitem [{\citenamefont {Dang}\ \emph
  {et~al.}(2026{\natexlab{b}})\citenamefont {Dang}, \citenamefont {Karur},\
  and\ \citenamefont {Sen}}]{Dang:2026xng}%
  \BibitemOpen
  \bibfield  {author} {\bibinfo {author} {\bibfnamefont {J.}~\bibnamefont
  {Dang}}, \bibinfo {author} {\bibfnamefont {R.}~\bibnamefont {Karur}},\ and\
  \bibinfo {author} {\bibfnamefont {S.}~\bibnamefont {Sen}},\ }\href@noop {}
  {\bibinfo {title} {{Gauge field flow for chiral gauge theories on a slab}}}
  (\bibinfo {year} {2026}{\natexlab{b}}),\ \Eprint
  {https://arxiv.org/abs/2606.05306} {arXiv:2606.05306 [hep-lat]} \BibitemShut
  {NoStop}%
\bibitem [{\citenamefont {Kroth}\ and\ \citenamefont
  {Sen}(2025)}]{Kroth:2025hba}%
  \BibitemOpen
  \bibfield  {author} {\bibinfo {author} {\bibfnamefont {J.~D.}\ \bibnamefont
  {Kroth}}\ and\ \bibinfo {author} {\bibfnamefont {S.}~\bibnamefont {Sen}},\
  }\bibfield  {title} {\bibinfo {title} {{Unpaired Weyl fermion on an axion
  string in a finite lattice}},\ }\href {https://doi.org/10.1103/twkg-8p31}
  {\bibfield  {journal} {\bibinfo  {journal} {Phys. Rev. D}\ }\textbf {\bibinfo
  {volume} {112}},\ \bibinfo {pages} {034510} (\bibinfo {year} {2025})},\
  \Eprint {https://arxiv.org/abs/2506.04324} {arXiv:2506.04324 [hep-th]}
  \BibitemShut {NoStop}%
\bibitem [{\citenamefont {Araki}\ \emph {et~al.}(2026)\citenamefont {Araki},
  \citenamefont {Fukaya}, \citenamefont {Onogi},\ and\ \citenamefont
  {Yamaguchi}}]{Araki:2026mqi}%
  \BibitemOpen
  \bibfield  {author} {\bibinfo {author} {\bibfnamefont {S.}~\bibnamefont
  {Araki}}, \bibinfo {author} {\bibfnamefont {H.}~\bibnamefont {Fukaya}},
  \bibinfo {author} {\bibfnamefont {T.}~\bibnamefont {Onogi}},\ and\ \bibinfo
  {author} {\bibfnamefont {S.}~\bibnamefont {Yamaguchi}},\ }\href@noop {}
  {\bibinfo {title} {{Symmetric Mass Generation for Domain-Wall Fermions}}}
  (\bibinfo {year} {2026}),\ \Eprint {https://arxiv.org/abs/2608.29963}
  {arXiv:2608.29963 [hep-lat]} \BibitemShut {NoStop}%
\bibitem [{\citenamefont {Aoki}\ \emph {et~al.}(2024)\citenamefont {Aoki},
  \citenamefont {Fukaya},\ and\ \citenamefont {Kan}}]{Aoki:2024bwx}%
  \BibitemOpen
  \bibfield  {author} {\bibinfo {author} {\bibfnamefont {S.}~\bibnamefont
  {Aoki}}, \bibinfo {author} {\bibfnamefont {H.}~\bibnamefont {Fukaya}},\ and\
  \bibinfo {author} {\bibfnamefont {N.}~\bibnamefont {Kan}},\ }\bibfield
  {title} {\bibinfo {title} {{A Lattice Formulation of Weyl Fermions on a
  Single Curved Surface}},\ }\href {https://doi.org/10.1093/ptep/ptae041}
  {\bibfield  {journal} {\bibinfo  {journal} {PTEP}\ }\textbf {\bibinfo
  {volume} {2024}},\ \bibinfo {pages} {043B05} (\bibinfo {year} {2024})},\
  \Eprint {https://arxiv.org/abs/2402.09774} {arXiv:2402.09774 [hep-lat]}
  \BibitemShut {NoStop}%
\bibitem [{\citenamefont {Aoki}\ and\ \citenamefont
  {Fukaya}(2022)}]{Aoki:2022cwg}%
  \BibitemOpen
  \bibfield  {author} {\bibinfo {author} {\bibfnamefont {S.}~\bibnamefont
  {Aoki}}\ and\ \bibinfo {author} {\bibfnamefont {H.}~\bibnamefont {Fukaya}},\
  }\bibfield  {title} {\bibinfo {title} {{Curved domain-wall fermions}},\
  }\href {https://doi.org/10.1093/ptep/ptac075} {\bibfield  {journal} {\bibinfo
   {journal} {PTEP}\ }\textbf {\bibinfo {volume} {2022}},\ \bibinfo {pages}
  {063B04} (\bibinfo {year} {2022})},\ \Eprint
  {https://arxiv.org/abs/2203.03782} {arXiv:2203.03782 [hep-lat]} \BibitemShut
  {NoStop}%
\bibitem [{\citenamefont {Aoki}\ and\ \citenamefont
  {Fukaya}(2023)}]{Aoki:2022aez}%
  \BibitemOpen
  \bibfield  {author} {\bibinfo {author} {\bibfnamefont {S.}~\bibnamefont
  {Aoki}}\ and\ \bibinfo {author} {\bibfnamefont {H.}~\bibnamefont {Fukaya}},\
  }\bibfield  {title} {\bibinfo {title} {{Curved domain-wall fermion and its
  anomaly inflow}},\ }\href {https://doi.org/10.1093/ptep/ptad023} {\bibfield
  {journal} {\bibinfo  {journal} {PTEP}\ }\textbf {\bibinfo {volume} {2023}},\
  \bibinfo {pages} {033B05} (\bibinfo {year} {2023})},\ \Eprint
  {https://arxiv.org/abs/2212.11583} {arXiv:2212.11583 [hep-lat]} \BibitemShut
  {NoStop}%
\bibitem [{\citenamefont {Karsten}(1981)}]{Karsten:1981gd}%
  \BibitemOpen
  \bibfield  {author} {\bibinfo {author} {\bibfnamefont {L.~H.}\ \bibnamefont
  {Karsten}},\ }\bibfield  {title} {\bibinfo {title} {{Lattice Fermions in
  Euclidean Space-time}},\ }\href
  {https://doi.org/10.1016/0370-2693(81)90133-7} {\bibfield  {journal}
  {\bibinfo  {journal} {Phys. Lett. B}\ }\textbf {\bibinfo {volume} {104}},\
  \bibinfo {pages} {315} (\bibinfo {year} {1981})}\BibitemShut {NoStop}%
\bibitem [{\citenamefont {Nielsen}\ and\ \citenamefont
  {Ninomiya}(1981{\natexlab{a}})}]{Nielsen:1980rz}%
  \BibitemOpen
  \bibfield  {author} {\bibinfo {author} {\bibfnamefont {H.~B.}\ \bibnamefont
  {Nielsen}}\ and\ \bibinfo {author} {\bibfnamefont {M.}~\bibnamefont
  {Ninomiya}},\ }\bibfield  {title} {\bibinfo {title} {{Absence of Neutrinos on
  a Lattice. 1. Proof by Homotopy Theory}},\ }\href
  {https://doi.org/10.1016/0550-3213(82)90011-6} {\bibfield  {journal}
  {\bibinfo  {journal} {Nucl. Phys. B}\ }\textbf {\bibinfo {volume} {185}},\
  \bibinfo {pages} {20} (\bibinfo {year} {1981}{\natexlab{a}})},\ \bibinfo
  {note} {[Erratum: Nucl.Phys.B 195, 541 (1982)]}\BibitemShut {NoStop}%
\bibitem [{\citenamefont {Nielsen}\ and\ \citenamefont
  {Ninomiya}(1981{\natexlab{b}})}]{Nielsen:1981hk}%
  \BibitemOpen
  \bibfield  {author} {\bibinfo {author} {\bibfnamefont {H.~B.}\ \bibnamefont
  {Nielsen}}\ and\ \bibinfo {author} {\bibfnamefont {M.}~\bibnamefont
  {Ninomiya}},\ }\bibfield  {title} {\bibinfo {title} {{No Go Theorem for
  Regularizing Chiral Fermions}},\ }\href
  {https://doi.org/10.1016/0370-2693(81)91026-1} {\bibfield  {journal}
  {\bibinfo  {journal} {Phys. Lett. B}\ }\textbf {\bibinfo {volume} {105}},\
  \bibinfo {pages} {219} (\bibinfo {year} {1981}{\natexlab{b}})}\BibitemShut
  {NoStop}%
\bibitem [{\citenamefont {Nielsen}\ and\ \citenamefont
  {Ninomiya}(1981{\natexlab{c}})}]{Nielsen:1981xu}%
  \BibitemOpen
  \bibfield  {author} {\bibinfo {author} {\bibfnamefont {H.~B.}\ \bibnamefont
  {Nielsen}}\ and\ \bibinfo {author} {\bibfnamefont {M.}~\bibnamefont
  {Ninomiya}},\ }\bibfield  {title} {\bibinfo {title} {{Absence of Neutrinos on
  a Lattice. 2. Intuitive Topological Proof}},\ }\href
  {https://doi.org/10.1016/0550-3213(81)90524-1} {\bibfield  {journal}
  {\bibinfo  {journal} {Nucl. Phys. B}\ }\textbf {\bibinfo {volume} {193}},\
  \bibinfo {pages} {173} (\bibinfo {year} {1981}{\natexlab{c}})}\BibitemShut
  {NoStop}%
\bibitem [{\citenamefont {Craig}(2026)}]{Craig:2026soq}%
  \BibitemOpen
  \bibfield  {author} {\bibinfo {author} {\bibfnamefont {N.}~\bibnamefont
  {Craig}},\ }\bibfield  {title} {\bibinfo {title} {{A gauge-invariant measure
  for lattice chiral fermions}},\ }\href@noop {} {\  (\bibinfo {year}
  {2026})},\ \Eprint {https://arxiv.org/abs/2610.09093} {arXiv:2610.09093
  [hep-lat]} \BibitemShut {NoStop}%
\bibitem [{\citenamefont {Kaplan}\ and\ \citenamefont
  {Sen}(2026)}]{Kaplan:2024ezz}%
  \BibitemOpen
  \bibfield  {author} {\bibinfo {author} {\bibfnamefont {D.~B.}\ \bibnamefont
  {Kaplan}}\ and\ \bibinfo {author} {\bibfnamefont {S.}~\bibnamefont {Sen}},\
  }\bibfield  {title} {\bibinfo {title} {{Regulated chiral gauge theory and the
  strong CP problem}},\ }\href {https://doi.org/10.1103/wld4-f12c} {\bibfield
  {journal} {\bibinfo  {journal} {Phys. Rev. D}\ }\textbf {\bibinfo {volume}
  {114}},\ \bibinfo {pages} {014508} (\bibinfo {year} {2026})},\ \Eprint
  {https://arxiv.org/abs/2412.02024} {arXiv:2412.02024 [hep-lat]} \BibitemShut
  {NoStop}%
\bibitem [{\citenamefont {Singh}(2025)}]{Singh:2025sye}%
  \BibitemOpen
  \bibfield  {author} {\bibinfo {author} {\bibfnamefont {H.}~\bibnamefont
  {Singh}},\ }\href@noop {} {\bibinfo {title} {{Ginsparg-Wilson Hamiltonians
  with Improved Chiral Symmetry}}} (\bibinfo {year} {2025}),\ \Eprint
  {https://arxiv.org/abs/2505.20419} {arXiv:2505.20419 [hep-lat]} \BibitemShut
  {NoStop}%
\bibitem [{\citenamefont {Golterman}\ and\ \citenamefont
  {Shamir}(2024)}]{golterman2024conserved}%
  \BibitemOpen
  \bibfield  {author} {\bibinfo {author} {\bibfnamefont {M.}~\bibnamefont
  {Golterman}}\ and\ \bibinfo {author} {\bibfnamefont {Y.}~\bibnamefont
  {Shamir}},\ }\bibfield  {title} {\bibinfo {title} {Conserved currents in
  five-dimensional proposals for lattice chiral gauge theories},\ }\href
  {https://doi.org/10.1103/PhysRevD.109.114519} {\bibfield  {journal} {\bibinfo
   {journal} {Phys. Rev. D}\ }\textbf {\bibinfo {volume} {109}},\ \bibinfo
  {pages} {114519} (\bibinfo {year} {2024})},\ \Eprint
  {https://arxiv.org/abs/2404.16372} {arXiv:2404.16372} \BibitemShut {NoStop}%
\bibitem [{\citenamefont {'t~Hooft}(1976)}]{t1976symmetry}%
  \BibitemOpen
  \bibfield  {author} {\bibinfo {author} {\bibfnamefont {G.}~\bibnamefont
  {'t~Hooft}},\ }\bibfield  {title} {\bibinfo {title} {Symmetry breaking
  through {B}ell-{J}ackiw anomalies},\ }\href
  {https://doi.org/10.1103/PhysRevLett.37.8} {\bibfield  {journal} {\bibinfo
  {journal} {Physical Review Letters}\ }\textbf {\bibinfo {volume} {37}},\
  \bibinfo {pages} {8} (\bibinfo {year} {1976})}\BibitemShut {NoStop}%
\bibitem [{\citenamefont {L{\"u}scher}(2010)}]{Luscher:2010iy}%
  \BibitemOpen
  \bibfield  {author} {\bibinfo {author} {\bibfnamefont {M.}~\bibnamefont
  {L{\"u}scher}},\ }\bibfield  {title} {\bibinfo {title} {{Properties and uses
  of the Wilson flow in lattice QCD}},\ }\href
  {https://doi.org/10.1007/JHEP08(2010)071} {\bibfield  {journal} {\bibinfo
  {journal} {JHEP}\ }\textbf {\bibinfo {volume} {08}},\ \bibinfo {pages}
  {071}},\ \bibinfo {note} {[Erratum: JHEP 03, 092 (2014)]},\ \Eprint
  {https://arxiv.org/abs/1006.4518} {arXiv:1006.4518 [hep-lat]} \BibitemShut
  {NoStop}%
\bibitem [{Note1()}]{Note1}%
  \BibitemOpen
  \bibinfo {note} {This also holds in $n=1$.}\BibitemShut {Stop}%
\bibitem [{Note2()}]{Note2}%
  \BibitemOpen
  \bibinfo {note} {Note that the above constraint on the interior gauge field
  can easily be expressed in the form of a completely local action formulation
  where the constraint is implemented using a Lagrange multiplier field at
  every interior space-time points. The novelty is that such a local action
  leads to the precise non-local action in Eq.\ (\ref {eq:Leta}) that allows
  for the mostly gapped spectrum of $\eta '$ and can lead to the elimination of
  the strong CP problem from this construction.}\BibitemShut {Stop}%
\end{thebibliography}%
\end{document}